\documentclass[prb,preprint,superscriptaddress,letterpaper,citeautoscript]{revtex4-2}

\usepackage{graphicx}
\usepackage{bm}
\usepackage{amssymb}
\usepackage{color}
\usepackage{amsmath}

\usepackage[hypertexnames=false,colorlinks=true,linkcolor=blue,anchorcolor=black,citecolor=blue,filecolor=black,menucolor=black,runcolor=black,urlcolor=blue]{hyperref}

\newcommand{\ignore}[1]{}

\begin{document}

\title{Dirac mass induced by optical gain and loss}

\author{Letian Yu}
\thanks{These authors contributed equally to this work.}
\affiliation{Division of Physics and Applied Physics, School of Physical and Mathematical Sciences, Nanyang Technological University, Singapore 637371, Singapore}

\author{Haoran Xue}
\thanks{These authors contributed equally to this work.}
\affiliation{Department of Physics, The Chinese University of Hong Kong, Shatin, Hong Kong SAR, China}

\author{Ruixiang Guo}
\affiliation{Division of Physics and Applied Physics, School of Physical and Mathematical Sciences, Nanyang Technological University, Singapore 637371, Singapore}
\affiliation{Centre for Disruptive Photonic Technologies, Nanyang Technological University, Singapore 637371, Singapore}

\author{Eng Aik Chan}
\affiliation{Division of Physics and Applied Physics, School of Physical and Mathematical Sciences, Nanyang Technological University, Singapore 637371, Singapore}
\affiliation{Centre for Disruptive Photonic Technologies, Nanyang Technological University, Singapore 637371, Singapore}

\author{Yun Yong Terh}
\affiliation{Division of Physics and Applied Physics, School of Physical and Mathematical Sciences, Nanyang Technological University, Singapore 637371, Singapore}

\author{Cesare Soci}
\email{csoci@ntu.edu.sg}
\affiliation{Division of Physics and Applied Physics, School of Physical and Mathematical Sciences, Nanyang Technological University, Singapore 637371, Singapore}
\affiliation{Centre for Disruptive Photonic Technologies, Nanyang Technological University, Singapore 637371, Singapore}

\author{Baile Zhang}
\email{blzhang@ntu.edu.sg}
\affiliation{Division of Physics and Applied Physics, School of Physical and Mathematical Sciences, Nanyang Technological University, Singapore 637371, Singapore}
\affiliation{Centre for Disruptive Photonic Technologies, Nanyang Technological University, Singapore 637371, Singapore}

\author{Y. D. Chong}
\email{yidong@ntu.edu.sg}
\affiliation{Division of Physics and Applied Physics, School of Physical and Mathematical Sciences, Nanyang Technological University, Singapore 637371, Singapore}
\affiliation{Centre for Disruptive Photonic Technologies, Nanyang Technological University, Singapore 637371, Singapore}

\maketitle

\textbf{Mass is commonly regarded as an intrinsic property of matter, but modern physics reveals particle masses to have complex origins \cite{Wilczek2012}, such as the Higgs mechanism in high-energy physics \cite{Anderson1963, Higgs1964}. In crystal lattices such as graphene, relativistic Dirac particles can exist as low-energy quasiparticles \cite{CastroNeto2009} with masses imparted by lattice symmetry-breaking perturbations \cite{haldane1988model, Kane2005, Hasan2010, Qi2011}. These mass-generating mechanisms all assume Hermiticity, or the conservation of energy in detail. Using a photonic synthetic lattice, we show experimentally that Dirac masses can be generated via non-Hermitian perturbations based on optical gain and loss.  We then explore how the space-time engineering of the gain/loss-induced Dirac mass affects the quasiparticles. As we show, the quasiparticles undergo Klein tunnelling at spatial boundaries, but a local breaking of a non-Hermitian symmetry can produce a novel flux nonconservation effect at the domain walls. At a temporal boundary that abruptly flips the sign of the Dirac mass, we observe a variant of the time reflection phenomenon: in the nonrelativistic limit, the Dirac quasiparticle reverses its velocity, while in the relativistic limit the original velocity is retained.}

The process by which relativistic particles acquire mass is a topic of great relevance to multiple areas of physics.  In elementary particle physics, researchers are still investigating the details of how the Higgs field, discovered experimentally in 2012, imparts mass to fermions and bosons \cite{Anderson1963, Higgs1964, Aad2012}, and the mathematical derivation of the Yang-Mills mass gap is one of the unsolved Millennium Problems \cite{yangmills}. Meanwhile, in low-energy physics, the key features of relativistic particles, including mass, can manifest through the quasiparticles of materials such as graphene \cite{CastroNeto2009}. Adding a mass to these ``relativistic'' quasiparticles is equivalent to lifting a point degeneracy in the band structure \cite{haldane1988model, Kane2005, Hasan2010, Qi2011\ignore{Semenoff1984, Thonhauser2006}} and can be achieved through a number of different lattice perturbations, which are also associated with different topological phases of matter \cite{haldane1988model, BansilReview2016}.

Are there any hitherto unexplored mechanisms for adding mass to a relativistic particle or quasiparticle?  It is notable that previous mechanisms have all assumed Hermiticity, the symmetry responsible for the conservation of energy in detail.  However, in the fast-developing field of non-Hermitian and nonequilibrium systems \cite{Bergholtz2021, Zaletel2023\ignore{Liu2011, Zhang2012}}, one can deliberately design structures that break Hermiticity, such as by incorporating gain and/or loss (i.e., energy transfer with an unmonitored external system).  Certain non-Hermitian lattices, such as those possessing ``unbroken parity/time reversal symmetry,'' are known to host stable quasiparticles that do not blow up or decay over time \cite{Bender1998, Mostafazadeh2002}.  Unfortunately, Dirac quasiparticles, the simplest type of relativistic quasiparticle, are usually spoiled by gain and loss.  Previous studies have found that the introduction of gain/loss turns Dirac point degeneracies into pairs of exceptional points (EPs) or rings of EPs \cite{\ignore{Zhou2018}Szameit2011, Zhen2015, Cerjan2019}, which behave very differently from Hermitian degeneracies like Dirac points \cite{\ignore{Kato1995, Peng2014, Chang2014,ghosh2016ep,Zhang2018phonon, Zhao2019, Miri2019 Hokmabadi2019,zhang2019quantum, wang2020petermann,Wiersig2022}doppler2016dynamically, Oezdemir2019, Bergholtz2021}.  Band energies do not remain real everywhere in the vicinity of an EP, destabilizing quasiparticle wavefunctions \cite{Szameit2011, doppler2016dynamically\ignore{ghosh2016ep}}.  EPs also fuse bands together, necessitating alternative formulations of non-Hermitian band topology in which Dirac quasiparticles play no special role \cite{Bergholtz2021\ignore{liang2013topological, shen2018topological, qiang2023}}.

\begin{figure*}
  \centering
  \includegraphics[width=0.95\textwidth]{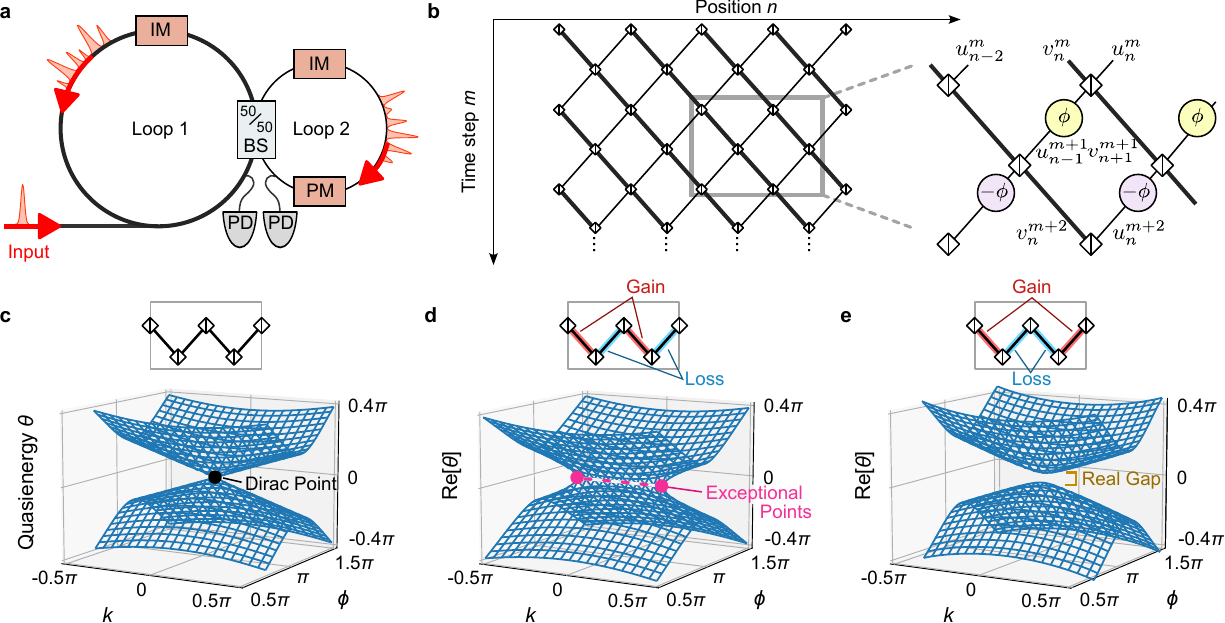}
  \caption{\textbf{Scheme for realizing a non-Hermitian synthetic lattice.}  \textbf{a}, Schematic of the experiment.  Two optical fiber loops are coupled by a 50/50 beamsplitter (BS).  Each fiber loop has an intensity modulator (IM), which can alter the amplitude of individual pulses. The short loop has an additional phase modulator (PM) to vary the phase of incoming signals. The pulses are measured by photodetectors (PDs) placed just before the BS.  \textbf{b}, The synthetic lattice that the pulse train evolution maps onto.  The index $n$ in each pulse train maps to a spatial position, and the number of round trips $m$ maps to time.  Inset: definitions of wavefunction amplitudes in the synthetic lattice.  \textbf{c}--\textbf{e}, Schematic of the distribution of gain/loss in the synthetic lattice (top), and the resulting quasienergy spectrum (bottom).  The Hermitian case (\textbf{c}) exhibits a Dirac point at $k = 0$, $\phi = \pi$.  Gain/loss can turn this into a pair of exceptional points (\textbf{d}) or a real mass gap (\textbf{e}).  In \textbf{d} and \textbf{e}, the gain/loss level is $g = 0.41$.  Only the two bands closest to quasienergy $\theta = 0$ are plotted. }
  \label{fig1}
\end{figure*}

In this paper, we experimentally demonstrate that gain and loss can be used to create stable massive relativistic quasiparticles.  We implement a looped optical fiber apparatus hosting a synthetic lattice \cite{Regensburger2012\ignore{Wimmer2015, Wimmer2017, Weidemann2020, Weidemann2022}} whose bandstructure has a Dirac point in the Hermitian limit.  Applying gain and loss, according to a specific ``semi-Hermitian'' symmetry \cite{Xue2020, Terh2023\ignore{Li2022,Li2022a}}, opens a relativistic mass gap (see Supplementary Information).  While the overall Hamiltonian is non-Hermitian, the two participating bands obey an effective Dirac Hamiltonian with a \textit{real} relativistic mass, whose sign and magnitude can be controlled via the gain/loss level. Although it was previously known that certain non-Hermitian bands can exhibit real band energies (e.g., through unbroken parity/time reversal symmetry \cite{\ignore{Bender1999}Bender1998, Mostafazadeh2002}), the use of gain/loss to generate a real mass for a Dirac quasiparticle has never been demonstrated, to our knowledge.

By controlling the gain and loss in the synthetic lattice, we find a number of interesting quasiparticle behaviors at spatial and temporal boundaries.

First, we investigate Klein tunneling, a counterintuitive property of the Dirac equation that allows particles to pass through a potential barrier with height $V$ exceeding the particle's rest energy $Mc^2$ (where $M$ is the rest mass and $c$ is the speed of light) \cite{holstein1998, dombey1999}. Although the Klein tunneling of massless Dirac particles has recently been demonstrated in a phononic crystal \cite{Katsnelson2006, Jiang2020}, the phenomenon has yet to be studied experimentally for massive Dirac particles.  We find that the Dirac quasiparticles in our lattice indeed exhibit Klein tunneling consistent with the Hermitian theory (e.g., total flux is conserved), contingent on semi-Hermiticity being preserved at the boundary. On the other hand, certain boundary choices can violate this symmetry, resulting in an anomalous variant of Klein tunneling whereby flux conservation fails at the boundary despite the quasiparticles having real Dirac masses on each side \cite{Terh2023}.

Second, we investigate the consequences of modulating the Dirac mass in time, and specifically abruptly reversing of its sign.  We find that such a ``temporal boundary'' leads to an interesting form of time reflection \cite{Bacot2016, Zhou2020, Lustig:21, Moussa2023, Dong2023}, or time-reversal of the quasiparticle wavefunction.  Recently, time reflection has been demonstrated in a number of notable experiments on water waves, electric waves, cold atoms, and other systems not described by the Dirac equation.  In the present synthetic lattice, we show that a mass-flipping temporal boundary selectively induces time reflection when the massive Dirac quasiparticle is in the nonrelativistic regime, through a state transfer between the upper and lower bands, an effect similar to that of time reversal on a particle (see Supplementary Information).  In the relativistic regime, however, the quasiparticle is unaffected by the temporal boundary.

Our experiment is based on a pair of coupled looped optical fibers (Fig.~\ref{fig1}\textbf{a} and Extended Data Fig.~\ref{figs3}).  A single optical pulse is initially injected into Loop 1, which is connected to the shorter Loop 2 by a 50/50 beamsplitter.  The initial pulse evolves into two pulse trains (one in each loop), which undergo repeated splitting and recombination at the beamsplitter.  The evolution of the pulse train maps onto a discrete-time ``light walk'' of a particle in a 1D synthetic lattice (Fig.~\ref{fig1}\textbf{b}), as established in previous works \cite{Regensburger2012, Wimmer2017\ignore{Wimmer2015, Wimmer2017, Weidemann2020, Weidemann2022}}.  The discrete position along each pulse train maps to the discrete lattice position $n$; the number of round trips the pulses have undergone map to the discrete time step $m$; and for each $(n,m)$, the pulses in Loop 1 and 2 represent right- and left-moving wave amplitudes in the synthetic lattice, denoted by $v_n^m$ and $u_n^m$ respectively (Fig.~\ref{fig1}\textbf{b}, inset).  For further details of the experimental setup, see Methods.

\begin{figure*}
  \centering
  \includegraphics[width=\textwidth]{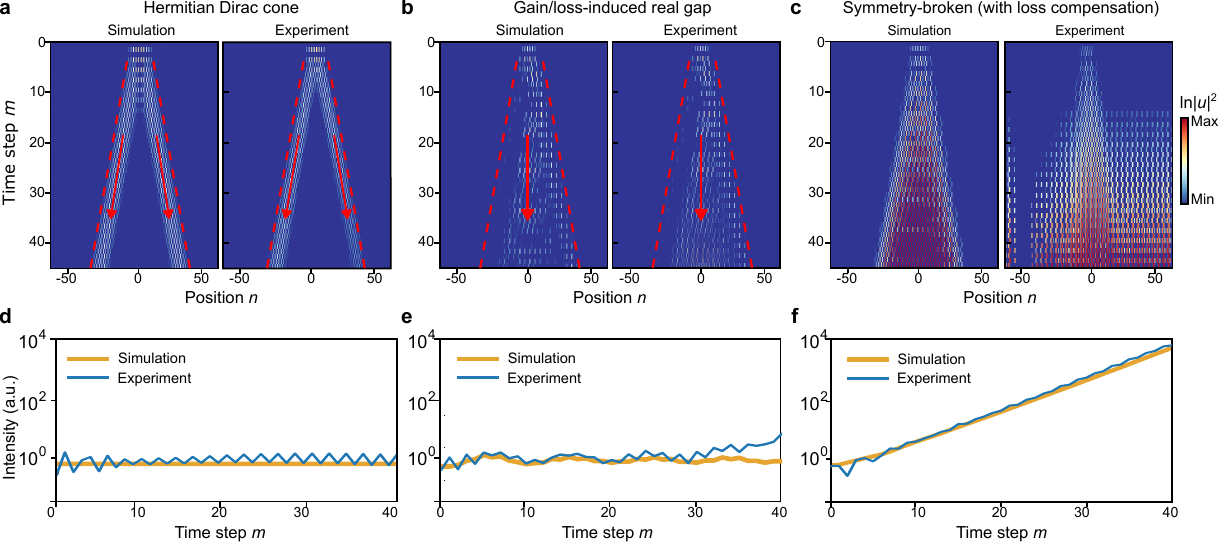}
  \caption{\textbf{Evolution of a gaussian wavepacket in the synthetic lattice.}  \textbf{a}, Evolution in the Hermitian lattice ($g = 0$), for an initial wavepacket centered at $k = 0$, $\phi = \pi$.  Conical diffraction (red arrows) occurs in both simulations (left panel) and experimental data (right panel).  \textbf{b}, Evolution in the non-Hermitian lattice ($g = 0.41$), corresponding to the massive relativistic quasiparticles of Fig.~\ref{fig1}\textbf{e}.  For the same initial wavepacket as in \textbf{a}, the wavefunction spreads throughout the light cone (red arrows).  Red dashes in \textbf{a}--\textbf{b} have slopes equal to the effective light speed $c = 1/\sqrt{2}$ (see Supplementary Information).  \textbf{c}, Similar to \textbf{b}, but with the initial wavepacket centered at $k = \phi = \pi$, where the band energies are non-real.  \textbf{d}--\textbf{f}, Total intensity of the pulse train versus time step $m$.  Conservation of total intensity in \textbf{e} shows that the non-Hermitian quasiparticles behave like Hermitian particles, unlike the exponential intensity increase seen in \textbf{f}.  In \textbf{c} and \textbf{f}, we perform the experiment with an additional loss on both fiber loops to avoid blow-up, and compensate for the loss when plotting.}
  \label{fig2}
\end{figure*}

Several electro-optical modulators, attached to the fiber loops, act as phase modulators (PMs) and intensity modulators (IMs) (Fig.~\ref{fig1}\textbf{a}).  The PM in Loop 2 applies a time-dependent phase modulation
\begin{equation}
  \Phi(m) = (-1)^m\phi,
\end{equation}
where $\phi$ is a tunable phase parameter.  The IMs, acting alongside fiber amplifiers \cite{Regensburger2012}, apply a spatially varying gain/loss factor $F(n)$ to each loop.  The resulting evolution equations are
\begin{align}
  \begin{aligned}
  u_n^{m+1}&=\frac{1}{\sqrt{2}}(u_{n+1}^{m}+iv_{n+1}^{m}) \, F(n) \, e^{i\Phi(m)},\\
  v_n^{m+1}&=\frac{1}{\sqrt{2}}(iu_{n-1}^{m}+v_{n-1}^{m}) \, F(n-1).
  \end{aligned}
  \label{eq:evolution}
\end{align}
For each $n$, we let $F(n)$ take two possible values,
\begin{equation}
  F(n) = \begin{cases} e^{g/2}, &\textrm{(gain)}\\
    e^{-g/2}, & \textrm{(loss)},

  \end{cases}
  \label{Fn}
\end{equation}
where $g$ parameterizes the gain/loss level.

Consider the Hermitian case, $g = 0$.  This lattice is translationally symmetric in both space and time, with period $\Delta n = \Delta m = 2$ (Fig.~\ref{fig1}\textbf{b}, inset).  We take a Floquet-Bloch ansatz \cite{Leykam2016}
\begin{equation}
  \begin{pmatrix}
    u_n^m \\ v_n^m 
  \end{pmatrix}
  =
  \begin{pmatrix}
    U \\ V
  \end{pmatrix}
  \exp\left[-\frac{im\theta}{\Delta m}\, + \frac{ikn}{\Delta n}\right],
  \label{ansatz}
\end{equation}
where $k$ is the quasimomentum and $\theta$ the quasienergy (both real).  Substituting into Eq.~\eqref{eq:evolution}, we obtain the Floquet bandstructure shown in Fig.~\ref{fig1}\textbf{c}, where $\theta$ is plotted against $k$ and the phase parameter $\phi$ (which serves as an additional parametric dimension \cite{Wimmer2017}).  It exhibits a Dirac point at $k = \theta = 0$, $\phi = \pi$.  (There is another Dirac point at $\theta = k= \pi$, $\phi = 0$, which we do not use.)  Here, we plot only the quasienergy bands near the Dirac point; a full band diagram is shown in Extended Data Fig.~\ref{figs1}.  

Next, we apply gain/loss to the lattice. We first consider a trivial case as shown in the upper panel of Fig.~\ref{fig1}\textbf{d}, the spatial period remains $\Delta n = 2$. The Dirac point turns into a pair of exceptional points (Fig.~\ref{fig1}\textbf{d}, lower panel), joined by a ``bulk Fermi arc'' along $\mathrm{Re}[\theta] = 0$, $\phi = \pi$ \cite{\ignore{Kozii2017, Zhou2018, GomisBresco2019}Oezdemir2018}.  Similar non-Hermitian bandstructures have been found in earlier works \cite{Zeuner2015\ignore{Su2021}}.  Near the exceptional line, the quasiparticles have $\mathrm{Im}[\theta] \ne 0$, and blow up or decay over time.

\begin{figure*}
  \centering
  \includegraphics[width=\textwidth]{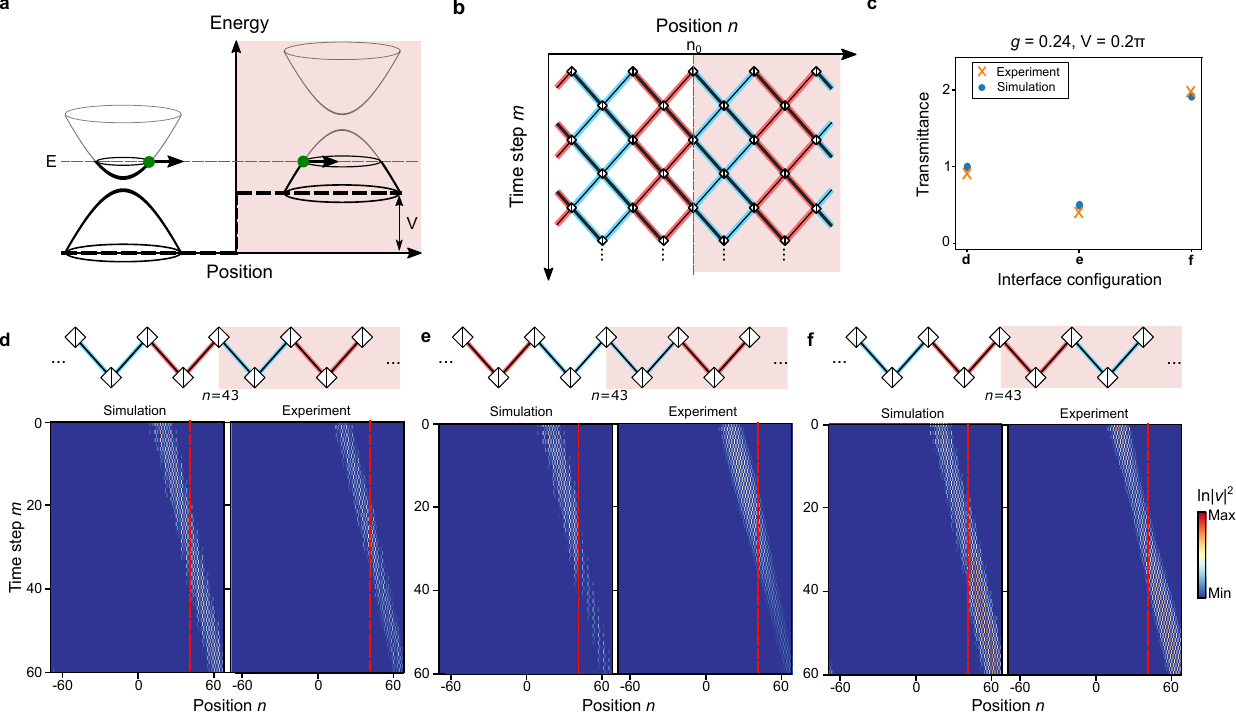}
  \caption{\textbf{Klein tunneling of massive Dirac quasiparticles.}  \textbf{a}, Schematic of Klein tunneling with massive Dirac cones. An incident quasiparticle of energy $E$ (left) impinges on a domain with scalar potential $V$ (right). \textbf{b}, Schematic of a synthetic lattice with a gain/loss distribution supporting a massive Dirac cone.  In the right domain, shaded in pink, a uniform scalar potential $V$ is applied.  \textbf{c}, Experimentally measured (cross) and simulated (dot) transmittance for the lattice configurations shown in \textbf{d}--\textbf{f}. \textbf{d}--\textbf{f}, Scattering results.  In all cases, the right domain has an additional scalar potential $V$, but the sites around the interface are laid out differently (top).  In the simulated (bottom left) and experimentally-measured (bottom right) intensity plots, reflection is found to be suppressed, while the transmittance is unity (\textbf{d}), damped (\textbf{e}), and amplified (\textbf{f}) respectively.  The lattice parameters are $\phi = \pi$, $g = 0.24$  (which corresponds to $M \approx 0.03)$, $k = 0.1\pi$  (which corresponds to $E \approx 0.02\pi$), and $V = 0.2\pi$. }
	\label{fig3}
\end{figure*}

Another gain/loss distribution with $\Delta n = 4$ is shown in the upper panel of Fig.~\ref{fig1}\textbf{e} (the temporal period remains $\Delta m = 2$). The Floquet-Bloch ansatz still has the form \eqref{ansatz}, but with a four-component wavefunction (see Supplementary Information). The spectrum now has a real quasienergy gap (Fig.~\ref{fig1}\textbf{e}, lower panel).  In the vicinity of $k = 0$, $\phi = \pi$, the two plotted bands are numerically verified to have exactly real $\theta$.  Moreover, they are governed by an effective Dirac Hamiltonian
\begin{equation}
  H_{\mathrm{eff}} = M(g) \sigma_1+\frac{1}{2}v_Dk\sigma_2 - v_D(\phi - \pi) \sigma_3,
  \label{2ddirac}
\end{equation}
where $\sigma_{1,2,3}$ are Pauli matrices, $v_D$ is the Dirac velocity, and $M(g) = \cosh g-1$ is a real Dirac mass that varies with the gain/loss level (see Extended Data Fig.~\ref{figs10} and Supplementary Information).  We also verified that the two participating bands are orthogonal (using the standard definition of the wavefunction inner product) for each $k$ and $\phi$.  These properties result from a ``semi-Hermitian'' symmetry which protects both the real-valuedness of eigenvalues and the pairwise orthogonality of eigenvectors, as shown in previous theoretical works \cite{Xue2020, Terh2023} (see Methods).

For Fig.~\ref{fig1}\textbf{e}, note that the bands are \textit{not} real everywhere: around $k = \pm \pi$, semi-Hermiticity is spontaneously broken \cite{Oezdemir2019} and the $\theta$'s are complex.

To study how quasiparticles evolve in the synthetic lattice, we use a diffusion protocol to prepare pulse trains corresponding to gaussian wavepackets centered around some $k$, with $\phi$ and other parameters fixed (see Methods).  We do not excite a specific band (which would involve specially preparing initial pulse sequences \cite{Wimmer2013}); instead, the initial wavepacket excites all Bloch states in all bands within a wavenumber window $\Delta k \approx 0.12\pi$.

For the Hermitian case ($g = 0$), with the initial wavepacket centered at $k = 0$, the wavefunction spreads in two well-defined beams (Fig.~\ref{fig2}\textbf{a}).  Simulation results obtained via Eq.~\eqref{eq:evolution} (left panel) agree well with the experimental data (right panel).  Such ``conical diffraction'' is a signature of a massless relativistic quasiparticle \cite{Peleg2007}.  The propagation speed is consistent with the effective light speed $c = 1/\sqrt{2}$ calculated from the Floquet bandstructure (see Supplementary Information).  Note that the $g = 0$ case has two distinct bands, which form a Dirac cone.  Fig.~\ref{fig2}\textbf{d} shows a logarithmic plot of the total intensity $I(m) = \sum_n \left(|u_n^m|^2 + |v_n^m|^2\right)$ versus $m$.  The graphs for both simulation and experiment are approximately constant, consistent with energy conservation.

Next, we apply the semi-Hermitian gain/loss distribution of Fig.~\ref{fig1}\textbf{e}, which is predicted to generate a real Dirac mass.  For the same initial conditions, the conical diffraction disppears and the wavefunction instead spreads over the entire light cone (Fig.~\ref{fig2}\textbf{b}).
These results are consistent with theoretical expectations: in the two participating bands, the quasiparticles become massive and thus possess a range of group velocities below the effective light speed.  There are also two other bands, whose quasienergies are real near $k = 0$ (Extended Data Fig.~\ref{figs1}\textbf{d}).  In Fig.~\ref{fig2}\textbf{e}, we see that the total intensity remains stable over long times, similar to the Hermitian case of Fig.~\ref{fig2}\textbf{d} and consistent with the Bloch states having real quasienergies.

For comparison, Fig.~\ref{fig2}\textbf{c} shows the evolution for a wavepacket centered at $k = \pi$, where the bands have non-real quasienergies.  All other parameters, including the initial conditions, are the same as in Fig.~\ref{fig2}\textbf{d}.  In the experiment, we apply an additional uniform loss to both fiber loops to prevent an intensity blow-up, and compensate for it when plotting (see Methods).  The results reveal exponential amplification over time (Fig.~\ref{fig2}\textbf{c},\,\textbf{f}) \cite{Miri2012a}.

\begin{figure*}
  \centering
  \includegraphics[width=\textwidth]{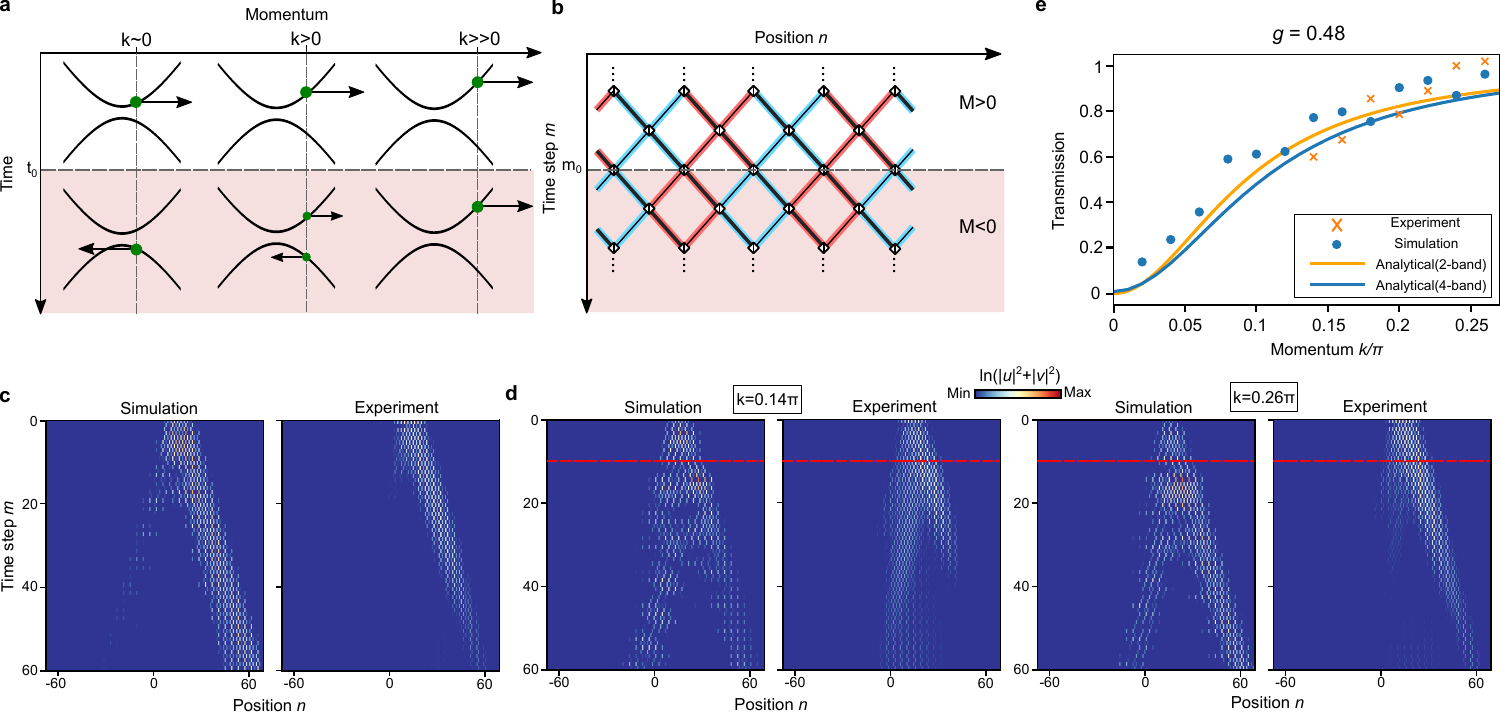}
  \caption{\textbf{Time reflection and refraction of Dirac quasiparticles.}  \textbf{a}, Schematic of time reflection and refraction with massive Dirac cones. An incident quasiparticle of energy $E$ (top) impinges on a temporal boundary that flips the sign of the Dirac mass (bottom). The value of the conserved momentum $k$ determines the scattering behavior.  \textbf{b}, Lattice configuration implementing the mass-flipping temporal boundary. \textbf{c}, Simulated and experimental intensity plots showing the free propagation of the wavepacket in the absence of the temporal boundary. \textbf{d}, Simulated and experimental intensity plots showing scattering from the mass-flipping temporal boundary (red dashes).  Reflection is observed at small $k$, and strong transmission at large $k$.  \textbf{e}, Transmittance versus $k$ based on experimental observations, lattice simulations, the non-Hermitian Dirac model (4 bands), and the effective Hermitian Dirac model (2 bands). The lattice parameters are $\phi = \pi$, $g = 0.48$ (which corresponds to $M \approx 0.12$), and $m_0 = 10$.}
	\label{fig4}
\end{figure*}

Next, we investigate the Klein tunneling of massive Dirac particles, as illustrated in Fig.~\ref{fig3}\textbf{a}. In the Hermitian case, Dirac particles can efficiently cross potential barriers of height $V$ exceeding $Mc^2$, with transmittance tending to unity as $V \rightarrow \infty$ \cite{holstein1998, dombey1999}. Semi-Hermitian Dirac quasiparticles, on the other hand, have been predicted to exhibit an anomalous form of Klein tunneling whereby energy conservation holds in the bulk but breaks down at the domain wall \cite{Terh2023} (this is unrelated to the claim of greater-than-unity reflectance in Hermitian Klein tunneling, which is an artifact of a redefinition of flux \cite{dombey1999}).  Here, we demonstrate both kinds of effects. As shown schematically in Fig.~\ref{fig3}\textbf{b}, we add an effective scalar potential barrier to the synthetic lattice (pink area) via a uniform phase modulation \cite{Ye2023} (see Methods).  There are several ways to configure the lattice boundary.  In the baseline case (Fig.~\ref{fig3}\textbf{d}), the wavepacket incident from the left experiences nearly unity transmission, with reflection strongly suppressed.  We also study alternative configurations where the gain/loss is flipped across the boundary, keeping the sign of $M$ unchanged.  In such cases, the reflection is still strongly suppressed, but the transmitted beam is either amplified (Fig.~\ref{fig3}\textbf{e}) or damped (Fig.~\ref{fig3}\textbf{e}).  This amplification or damping depends on the gain/loss of the boundary's central site, and occurs only while the wavepacket is impinging on the step.  Within the bulk domains, the intensity is conserved since $M$ is real.  Our experimental results for the net transmittance are in excellent agreement with simulations, as shown in Fig.~\ref{fig3}\textbf{c}. 
     
The scattering of Dirac particles by time-dependent modulations is also of significant physical interest.  In other metamaterial models, it has been shown that a wave experiencing an abrupt change in refractive index can undergo a splitting that is interpretable as time reflection and refraction \cite{Bacot2016, Zhou2020, Lustig:21, Moussa2023, Dong2023}.  Along similar lines, we consider the consquences of abruptly changing the mass, which is normally regarded as an intrinsic particle property.  Specifically, we implement a temporal boundary that abruptly flips the sign of the Dirac mass $M$ (Fig.~\ref{fig4}\textbf{a}).  Due to translational invariance, $k$ is conserved across the temporal boundary.  For stationary Dirac particles ($k=0$), the mass flip exactly interchanges the upper- and lower-band-edge eigenstates. In the nonrelativistic limit, time reflection should thus occur as the quasiparticle is efficiently transferred to a negative-energy state with the opposite group velocity (see Supplementary Information). Intriguingly, this behavior follows the simple intuition that given a conserved nonrelativistic momentum $p = Mv$, flipping the sign of $M$ causes $v$ to reverse. In the relativistic (large-$k$) limit, the mass-flip should be negligible, allowing the particle to be transmitted strongly across the temporal boundary.

To implement this temporal boundary, we apply the lattice configuration shown in Fig.~\ref{fig4}\textbf{b}. The gain/loss distributions are displaced by a quarter of a unit cell after $m_0=10$, reversing the sign of $M$ (see Supplementary Information). In the absence of the temporal boundary, a gaussian wavepacket prepared with momentum $k$ propagates with constant group velocity (Fig.~\ref{fig4}\textbf{c}). With the temporal boundary, we observe strong reflection at small $k$ (Fig.~\ref{fig4}\textbf{d}, left), and strong transmission at large $k$ (Fig.~\ref{fig4}\textbf{d}, right), consistent with the arguments in the previous paragraph.  Extracting the $k$-dependent transmission probabilities (using intensities at $m = 42$), we find a good quantitative match with the predictions of the non-Hermitian continuum Hamiltonian \eqref{2ddirac}, as well as the effective two-band Hermitian Dirac Hamiltonian, as shown in Fig.~\ref{fig4}\textbf{e}.

In conclusion, we have demonstrated a non-Hermitian mechanism for generating real Dirac masses, using a synthetic photonic lattice containing optical gain/loss and Floquet modulations.  Rather than being an intrinsic property, the Dirac mass can be engineered in space and time by varying the gain and loss, as we have demonstrated with our observations of non-Hermitian anomalous Klein tunneling and the selective time reflection of massive Dirac quasiparticles.  In the future, other exotic phenomena should be accessible with this experimental platform, such as non-Hermitian Landau levels \cite{Xue2020} or relativistic momentum gaps \cite{Lustig:21}.  Different varieties of relativistic quasiparticles may also be realizable using more complicated lattice designs \cite{\ignore{Goldman2014, Eckardt2017}Bukov2015}.

\vskip 0.1in
\noindent
{\large \textbf{Methods}}

\noindent
\textbf{Experimental setup}

\noindent
A schematic of the experiment is shown in Extended Data Fig.~\ref{figs3}.  A distributed feedback laser diode operating at $\lambda=1550\,\textrm{nm}$ sends a continuous-wave (CW) signal into our system. To prevent backscattering into the laser source, an isolator is placed immediately after it. The input signal is shaped into rectangular pulses of width $100\,\textrm{ns}$ by an electro-optical modulator (EOM). Subsequently, the signal is enhanced by an erbium-doped fiber amplifier (EDFA), and then cleaned up by a band pass filter (BPF) and a polarizer.

Before developing the signal into a synthetic lattice, we take additional steps to stabilize the amplification provided by the EDFA, based on the protocol reported in Ref.~\onlinecite{Regensburger2012}.  A quasi-continuous train of pulses, each separated by roughly one round trip time, is injected into both loops prior to the actual experiment. During this initial phase, the EOMs in both loops are set to 0\% transmission, allowing the EDFAs to adapt to the input power level. After about $8\,\textrm{ms}$, we set the EOM at the signal generation side to 0\%, so that only the last pulse from the warm-up phase is retained for the experiment.

After the warm-up phase, the pulses begin to propagate within the coupled fiber loops.  The average round trip time is about $20\,\mu\textrm{s}$, with a difference of about $300\,\textrm{ns}$ between the loops. To further stablize the pulses, a strong pilot signal at $1535\,\textrm{nm}$ (the wavelength corresponding to the peak of EDFA emission spectrum) is applied in both loops right after the fiber spools. This signal can be cleaned by a following BPF due to its large wavelength difference with our operating wavelength of $\lambda=1550\,\textrm{nm}$. Polarizers and polarizing beamsplitters are employed to ensure that only a single polarization is present in both loops. 

We remotely control the electro-optical modulator (EOM) and phase modulators (PM) to implement the gain/loss and phase modulation profiles described in the main text.  We also use these to implement absorbing boundary conditions for the synthetic lattice, by setting the transmission in both loops to 0\% at the start or near the end of each round trip.  These settings also help prevent noise build-up inside the loops.

To observe the wavefunction dynamics, we monitor the pulses circulating in each loop with photodiodes (PDs) placed before the central coupler.

\vskip 0.1in
\noindent
\textbf{Semi-Hermitian Floquet bandstructures}

\noindent
The non-Hermitian Floquet bandstructures we consider (configuration in Fig.~\ref{fig1}\textbf{e}) can host a massive Dirac Hamiltonian because of a pair of symmetries we call ``semi-Hermiticity'' \cite{Xue2020, Terh2023}.  In terms of their action on a $4\times4$ evolution operator $U$, the symmetries are:
\begin{align}
  \Sigma_0 U \Sigma_0 &= (U^{-1})^\dag \label{pseudoHerm} \\ 
  \{ iU,\Sigma_3\Sigma_1 T \}&=0, \label{antiPT}
\end{align}
where $T$ is the complex conjugation operator and
\begin{equation}
\Sigma_\mu = 
\begin{bmatrix}
0 & \sigma_\mu \\
\sigma_\mu & 0 
\end{bmatrix}
,\quad
\mu = 0,1,2,3.
\end{equation}
Here, $\sigma_{0}$ is the $2\times2$ identity matrix, and $\sigma_1$, $\sigma_2$, and $\sigma_3$ are the Pauli matrices.

We first consider Eq.~\eqref{pseudoHerm}, which is an instance of pseudo-Hermiticity \cite{Mostafazadeh2002}.  $U$ has a set of right eigenvectors, obtained from the eigenproblem
\begin{equation}
  U |\psi_n\rangle = e^{-i\theta_n} |\psi_n\rangle,
  \label{Ueig}
\end{equation}
where $\theta_n \in \mathbb{C}$ is the quasienergy.  The corresponding left eigenvectors satisfy
\begin{align}
  \langle\phi_n| U &= \langle\phi_n| e^{-i\theta_n}\\
  \Rightarrow \quad \left(U^{-1}\right)^\dagger |\phi_n\rangle
  &= e^{-i\theta_n^*} |\phi_n\rangle.
\end{align}
Therefore Eq.~\eqref{pseudoHerm} implies that
\begin{equation}
  U \Big(\Sigma_0|\phi_n\rangle\Big)
  = \Sigma_0 (U^{-1})^\dag |\phi_n\rangle
  = e^{-i\theta_n^*} \Big(\Sigma_0|\phi_n\rangle\Big).
\end{equation}
Thus, for each solution to Eq.~\eqref{Ueig} with quasienergy $\theta_n$, there is a solution with quasienergy $\theta_n^*$.  The quasienergies are either real, or come in complex conjugate pairs.

Eq.~\eqref{antiPT} is an instance of anti-PT symmetry.  It is equivalent to
\begin{equation}
  U^*\Sigma_3\Sigma_1 = \Sigma_3\Sigma_1 U.
\end{equation}
Using this with Eq.~\eqref{Ueig} gives
\begin{align}
  \Sigma_3\Sigma_1 U |\psi_n\rangle &= U^* \Sigma_3\Sigma_1 |\psi_n\rangle
  = e^{-i\theta_n} \Sigma_3\Sigma_1 |\psi_n\rangle \\
  UT \Sigma_3\Sigma_1 |\psi_n\rangle
  &= e^{i\theta_n^*} T \Sigma_3\Sigma_1 |\psi_n\rangle.
\end{align}
Hence, there is an eigenstate $T \Sigma_3\Sigma_1 |\psi_n\rangle$ with quasienergy $-\theta_n^*$.  Moreover, we can show that this is orthogonal to the original eigenstate, as follows: let
\begin{equation}
|\psi_n\rangle = 
\begin{pmatrix} p \\ q \end{pmatrix},
\end{equation}
where $p$ and $q$ are two-component vectors.  Then
\begin{align}
  \langle\psi_n | T\Sigma_3\Sigma_1 | \psi_n\rangle
  &= \begin{pmatrix} p^\dagger & q^\dagger
  \end{pmatrix}
  \begin{pmatrix}-i\sigma_2 & 0 \\ 0 & -i\sigma_2 \end{pmatrix}
  \begin{pmatrix}p^* \\ q^*
  \end{pmatrix} \\
  &= -i \left(p^\dagger \sigma_2 p^* + q^\dagger \sigma_2 q^*\right) \\
  &= 0 \;\;\; \textrm{for any $p$, $q$}.
\end{align}
Thus, the combination of Eqs.~\eqref{pseudoHerm} and \eqref{antiPT} can produce behavior analogous to Hermiticity.  If the pseudo-Hermiticity \eqref{pseudoHerm} is unbroken, then \eqref{antiPT} implies the existence of pairs of orthogonal eigenstates with real eigenvalues $(\theta_n, -\theta_n)$ where $\theta_n \in \mathbb{R}$ \cite{Xue2020}.

We thus see that the combination of two non-Hermitian symmetry conditions---unbroken pseudo-Hermiticity and anti-PT symmetry---yield the two features (real eigenvalues and pairwise-orthogonal eigenvectors) consistent with the existence of an effective Dirac Hamiltonian.  Satisfying both conditions simultaneously requires at least a $4\times 4$ matrix; hence, aside from the two orthogonal bands participating in the Dirac cone, there are two other auxiliary bands (which are likewise orthogonal to each other).

Note that a $2\times 2$ matrix cannot give this behavior.  In the $2\times 2$ case, having two real eigenvalues with orthogonal eigenvectors implies that the Hamiltonian is Hermitian (for a 2D vector space, pairwise orthogonality simply implies that the two eigenvectors form an orthogonal basis).

\vskip 0.1in
\noindent
\textbf{State preparation}

\noindent
For the experiments of Fig.~\ref{fig2}, we prepare initial wavepackets designed to excite quasiparticles around a target momentum $k$.  This is achieved using the procedure described in Ref.~\onlinecite{Wimmer2013}.  We start with a single pulse produced by the previously-described signal generation system.  This is allowed to evolve according to the evolution equations
\begin{align}
  u_n^{m+1}&=\frac{1}{\sqrt{2}} \left(u_{n+1}^{m}+iv_{n+1}^{m}\right) \, e^{i\phi_0}\, F(m)
  \label{prep1} \\
  v_n^{m+1}&=\frac{1}{\sqrt{2}} \left(iu_{n-1}^{m}+v_{n-1}^{m}\right),
\end{align}
where
\begin{equation}
  F(m) = m \; (\textrm{mod}~2).
\end{equation}
This is equivalent to fully absorbing the signal inside the short loop in every two time steps.  To select the Bloch modes at wavenumber $k$, we choose the phase modulation $\phi_0$ such that
\begin{equation}
  k = 2\pi+2\phi_0.
  \label{prep4}
\end{equation}
Extended Data Fig.~\ref{figs4}\textbf{a} shows the evolution of the pulses, and Extended Data Fig.~\ref{figs4}\textbf{b} shows the intensity distribution at the end of the preparation stage ($m = 55$).  The experimental results are clearly in excellent agreement with simulations, as well as fitting a Gaussian envelope
\begin{equation}
  f(n) = A\exp\left[-\frac{(n-n_0)^2}{2\sigma^2}\right],
\end{equation}  
where $n_0$ is the center of the wavepacket and $\sigma$ is its spatial width.  In our case, $\sigma\approx3.7$, corresponding to a spectral width of $\delta\theta \approx 0.12\pi$.

\vskip 0.1in
\noindent
\textbf{Probing states with complex quasienergies}

\noindent
In Fig.~\ref{fig2}\textbf{c},\textbf{f}, we plot the time evolution of a wavepacket centered at $k = \pi$, where the band quasienergies are non-real.  In contrast to the $k = 0$ case, where the quasiparticles behave like energy-conserving Dirac particles, in this case the wavefunction is predicted to undergo exponential amplification.

Experimentally, it is not desirable to access the amplifying regime, as the amplification of the EDFAs will saturate \cite{Regensburger2012}.  To bypass this problem, we employ a simple compensation method previously used, e.g., in Ref.~\onlinecite{Weidemann2022}.  We apply an additional loss, which is uniform in both space and time, to both loops, at a loss level that maintains nearly constant overall power throughout the experiment (based on the PD observations).  The evolution equations are modified to
\begin{align}
  u_n^{m+1}&=\frac{1}{\sqrt{2}}(u_{n+1}^{m}+iv_{n+1}^{m})e^{i\Phi(m)}F(n) e^{-\gamma/2}\\
  v_n^{m+1}&=\frac{1}{\sqrt{2}}(iu_{n-1}^{m}+v_{n-1}^{m})F(n-1) e^{-\gamma/2},
\end{align}
where $\exp(-\gamma/2)$ is the damping factor over one time step.  The uniform loss does not alter the bandstructure, apart from adding a constant imaginary part to the quasienergy of every eigenstate, equivalent to a spatially uniform decay factor $\exp(-\gamma m/2)$.

In Extended Data Fig.~\ref{figs16}, we plot the Fourier spectra obtained from the numerically simulated evolving wavefunctions, for the three different unit cell configurations of Fig.~\ref{fig1}\textbf{c}--\textbf{e} with $\phi = \pi$. The results show good agreement with the theoretical Floquet bandstructures (the imaginary parts of the bands manifest as bright regions in the Fourier spectra).  Experimentally accessing the corresponding phase information is impractical with our current system parameters; with a 20 $\mu\textrm{s}$ average round trip time and 8 ms warm-up interval, a reference field with coherence length of up to 200 km would be required.

In Extended Data Fig.~\ref{figs6}\textbf{a}, we plot the raw (uncompensated) time evolution of a wavepacket subject to this uniform loss.  The total intensity does not undergo exponential amplification, as shown in Extended Data Fig.~\ref{figs6}\textbf{b}.  We then multiply the wavefunctions by $\exp(\gamma m/2)$ to produce the experimental data shown in Fig.~\ref{fig2}.

%
%

\vskip 0.1in
\noindent
\textbf{Topological nature of the Dirac mass}

\noindent
One of the most remarkable features of massive Dirac quasiparticles is that they act as ``building blocks'' for topological band insulators.  A massive 2D Dirac Hamiltonian, Eq.~\eqref{2ddirac}, contributes $\pm\textrm{sgn}(m)/2$ to the Chern numbers of the two participating bands \cite{haldane1988model\ignore{Shen2017}}.  When there are two adjacent domains with opposite signs of $m$, topological states appear at the boundary \cite{JackiwRebbi1976, Angelakis2014}.

To show that the gain/loss induced Dirac masses can similarly create topological boundary states, we implement a lattice consisting of two domains, whose gain/loss distributions are displaced relative to each other by a quarter of a unit cell (Extended Data Fig.~\ref{fig3supp}\textbf{a}).  From the derivation of the Floquet bandstructure, we can show that this displacement flips the sign of $M$ (see Supplementary Information).  (In an earlier theoretical analysis of a non-Hermitian 1D topological lattice, a similar displacement was also shown to induce a topological transition \cite{Takata2018}.)  In Extended Data Fig.~\ref{fig3supp}\textbf{b}, the quasienergy bands for a finite sample are plotted against $\phi$ (which serves as a momentum-like variable).  This reveals a branch of boundary states spanning the gain/loss induced band gap.  They have a chiral dispersion relation, as expected of the topological boundary states generated by a Dirac mass inversion, even though $\mathrm{Im}(\theta)$ does not strictly vanish because the domain wall breaks the semi-Hermitian symmetry \cite{Takata2018}.  From the spatial distributions (Extended Data Fig.~\ref{fig3supp}\textbf{c}), we see that the localization length is affected by the gain/loss level $g$, which governs the size of the bulk gap.  We emphasize that these topological boundary states are solely induced by gain/loss, since the bulk gap is closed for $g = 0$.

To probe this phenomenon experimentally, we inject a single initial pulse at the boundary ($n = 0$), which has strong spatial overlap with the boundary states.  In Extended Data Fig.~\ref{fig3supp}\textbf{d}, we plot the intensity distribution for $g = 0.41$ (blue circles), obtained by a normalized time average of $|u_n^m|^2+|v_n^m|^2$.  The profile fits the theoretically calculated spatial distribution for the boundary state (red line).  Finally, from the time evolution data for different values of $g$ (Extended Data Fig.~\ref{fig3supp}\textbf{e}--\textbf{g}), we observe that the localization length decreases with $g$, in agreement with Extended Data Fig.~\ref{fig3supp}\textbf{c}.

\vskip 0.1in
\noindent
\textbf{Klein tunneling setup}

\noindent
Here, we provide additional details about the Klein tunneling results shown in Fig.~\ref{fig3}. In these experiments, we use an initial wavepacket at $k=0.1\pi$ with a preparation stage of time step $m=105$. To form the barrier, we follow the method of Ref.~\cite{Ye2023}: instead of having a single phase modulator in the short loop, another phase modulator is added to the long loop. The evolution equation is then modified to  
\begin{align}
  \begin{aligned}
  u_n^{m+1}&=\frac{1}{\sqrt{2}}(u_{n+1}^{m}+iv_{n+1}^{m}) \, F(n) \, e^{i\Phi(m)-iV/2},\\
  v_n^{m+1}&=\frac{1}{\sqrt{2}}(iu_{n-1}^{m}+v_{n-1}^{m}) \, F(n-1)\, e^{-iV/2},
  \end{aligned}
  \label{eq:evolutionScatter}
\end{align}
where $V$ denotes the scalar potential (in Fig.~\ref{fig3}, we use $V=0.2\pi$).  The factor of $1/2$ is due to the Floquet unit cell having a time step of $2$.   The transmittance and intensity plots for other values of $V$ are shown in Extended Data Fig.~\ref{figs14}.

In Fig.~\ref{fig3}\textbf{c} and Extended Data Fig.~\ref{figs14}\textbf{b}, we plot the transmittance and reflectance, which respectively correspond to the intensities summed over sites to the right and left of the interface.  These intensities are evaluated at time step $m = 60$, after the scattering process has completed.  One complication is that during the wavepacket preparation stage, in addition to the right-propagating wavepacket (residing in the upper band), we also excite pulses of the opposite propagation direction (residing in the lower band), as shown in the short loop intensity evolution in Extended Data Fig.~\ref{figs14}\textbf{c}--\textbf{e}. These additional excitations do not participate in the scattering process, and are excluded when finding the reflectance and tramittance.

\medskip

\bibliography{references}

\vskip 0.1in
\noindent
\textbf{Acknowledgements} This work was supported by the Singapore Ministry of Education (MOE) Tier 1 Grant No. RG148/20, and by the National Research Foundation (NRF), Singapore under Competitive Research Programme NRF-CRP23-2019-0005 and NRF-CRP23-2019-0007, and NRF Investigatorship NRF-NRFI08-2022-0001. C.S. and R.G. acknowledge the support of the Quantum Engineering Programme of the Singapore National Research Foundation, grant number NRF2021-QEP2-01-P01.

\vskip 0.1in
\noindent
\textbf{Author Contributions} H.X., B.Z., and Y.D.C.~conceived the idea. H.X., L.Y., R.G., and E.A.C.~designed and performed the experiment. L.Y. and R.G.~analysed the data. C.S., B.Z., Y.D.C.~supervised the project. H.X., L.Y., R.G., Y.Y.T, C.S., B.Z., and Y.D.C.~contributed to the discussion of the results and writing of the manuscript.

\vskip 0.1in
\noindent
\textbf{Data availability} Correspondence and requests for materials should be addressed to Cesare Soci, Baile Zhang or Y.D. Chong.

\vskip 0.1in
\noindent
\textbf{Code availability} Correspondence and requests for materials should be addressed to Cesare Soci, Baile Zhang or Y.D. Chong.

\vskip 0.1in
\noindent
\textbf{Competing Interests} The authors declare no competing interests.

\vskip 0.1in
\noindent
\textbf{Additional Information} Correspondence and requests for materials should be addressed to Cesare Soci, Baile Zhang or Y.D. Chong.

\clearpage

\renewcommand{\figurename}{\textbf{Extended Data Figure}}
\setcounter{figure}{0}

\begin{figure*}
  \centering
  \includegraphics[width=\textwidth]{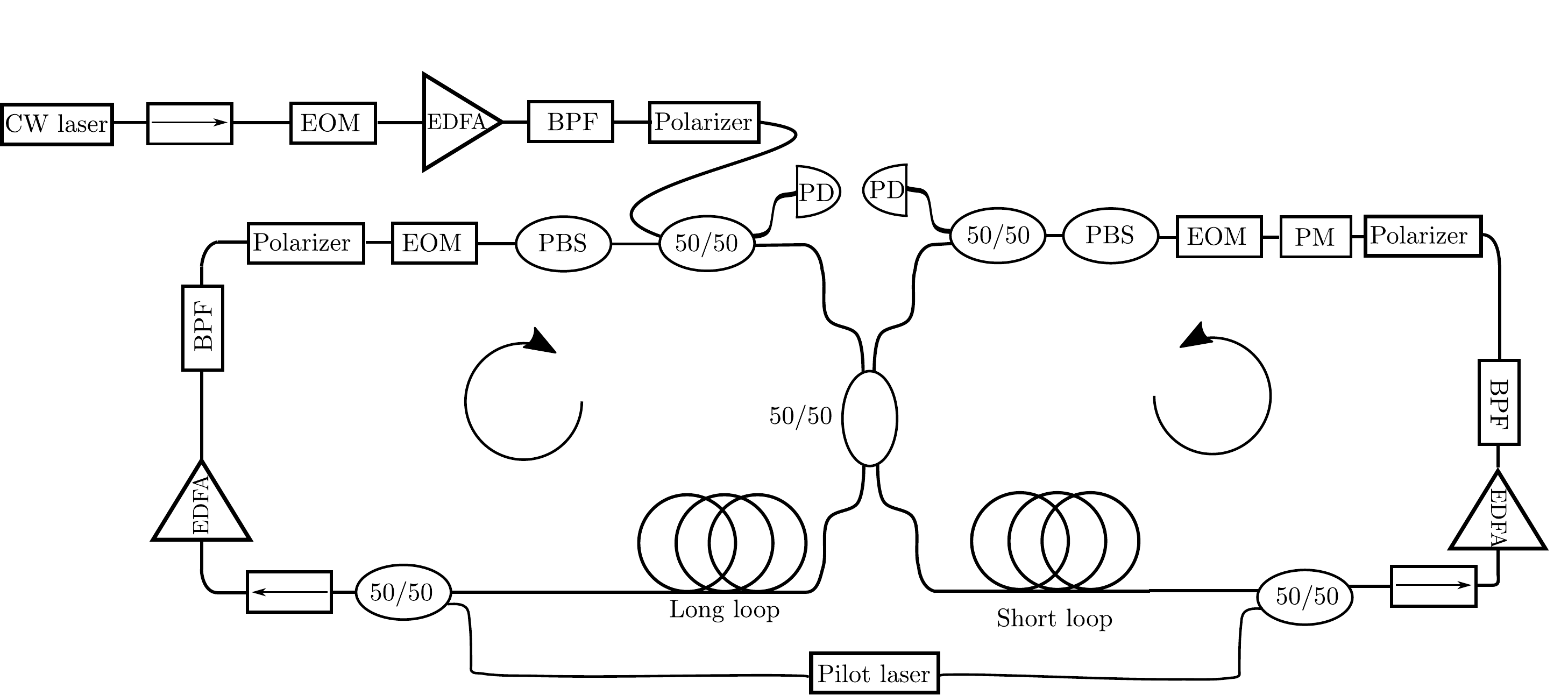}
  \caption{\textbf{Schematic of the experiment.}  The abbreviations are as follows: continuous-wave laser (CW), electro-optical modulator (EOM), erbium-doped fiber amplifier (EDFA), photodiode detector (PD), band-pass filter (BPF), polarization beam splitter (PBS), and phase modulator (PM). The boxes containing arrows denote isolators, and the ovals labeled `50/50' and `90/10' denote optical couplers with the indicated splitting ratios.}
  \label{figs3}
\end{figure*}

\begin{figure*}
  \centering
  \includegraphics[width=\textwidth]{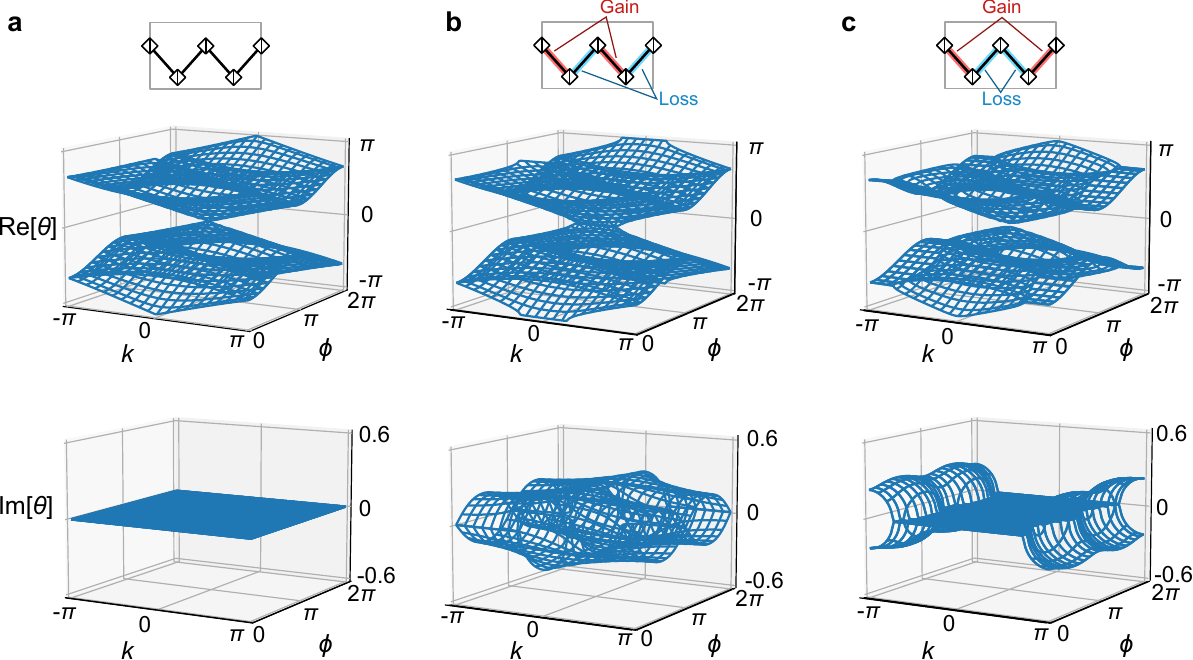}
  \caption{\textbf{Floquet band diagrams of the synthetic lattice for different gain/loss and phase distributions}.  \textbf{a}, No gain/loss ($g = 0$).  \textbf{b}, Gain/loss/gain/loss.  The Dirac point turns into a pair of EPs.  The band energies are non-real at $k = 0$, $\phi = \pi$, while the band energies close to $k =\pm \pi$, $\phi = \pi$ are real. \textbf{c}, Gain/loss/loss/gain.  At $k = 0$, $\phi = \pi$, there are two orthogonal bands with real band energies, governed by a massive Dirac Hamiltonian.  In \textbf{b}--\textbf{c}, the gain/loss level is $g = 0.41$.}
  \label{figs1}
\end{figure*}

\begin{figure*}
  \centering
  \includegraphics[width=0.6\textwidth]{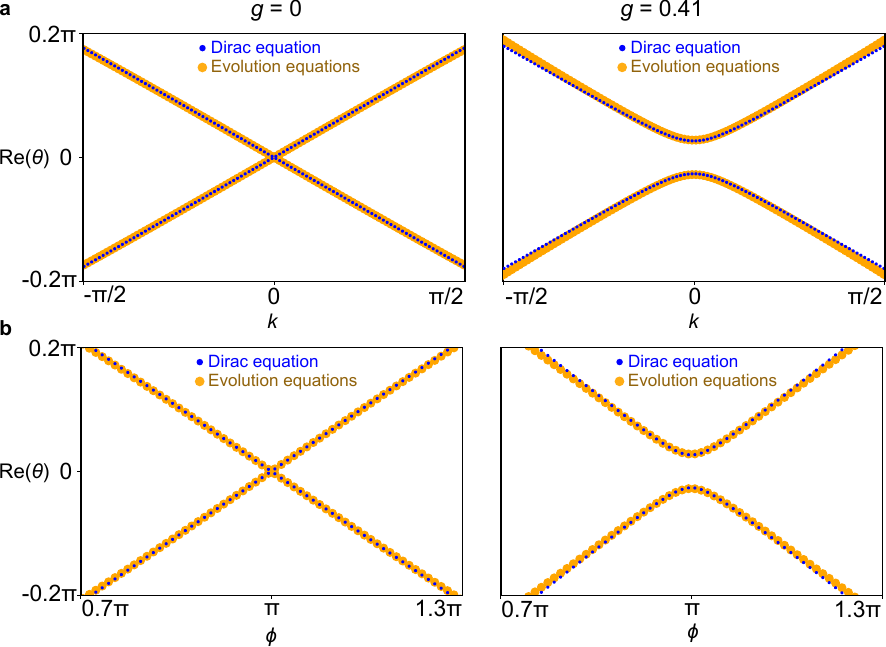}
  \caption{\textbf{Dispersion relation for the Dirac quasiparticles}.  These Floquet band diagrams are calculated using the evolution equations (orange) and the effective Dirac Hamiltonian (blue; see Supplementary Information), for $g = 0$ (left) and $g = 0.41$ (right).  \textbf{a}, Quasienergies versus $k$.  \textbf{b}, Quasienergies versus $\phi$.  }
  \label{figs10}
\end{figure*}

\begin{figure*}
  \centering
  \includegraphics[width=0.8\textwidth]{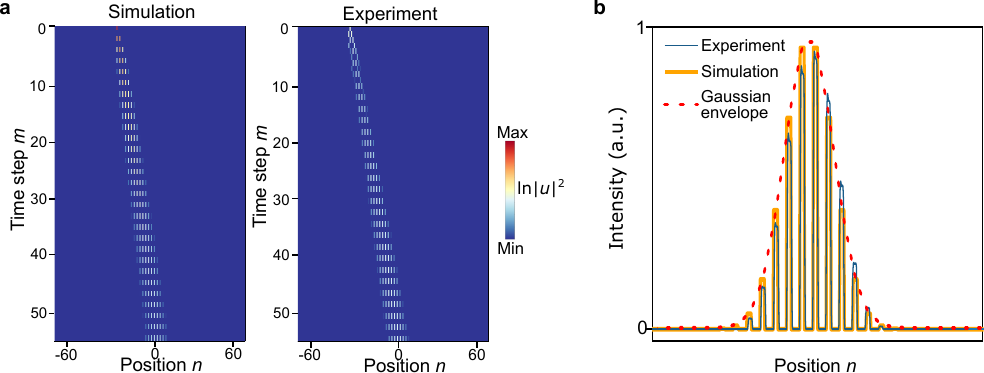}
  \caption{\textbf{Preparation of pulse train for exciting quasiparticles}. \textbf{a}, Time evolution of a single initial pulse under Eqs.~\eqref{prep1}--\eqref{prep4}, for $\phi_0 = -\pi$.  \textbf{b}, Intensity profile at $m = 55$.}
  \label{figs4}
\end{figure*}

\begin{figure*}
  \centering
  \includegraphics[width=\textwidth]{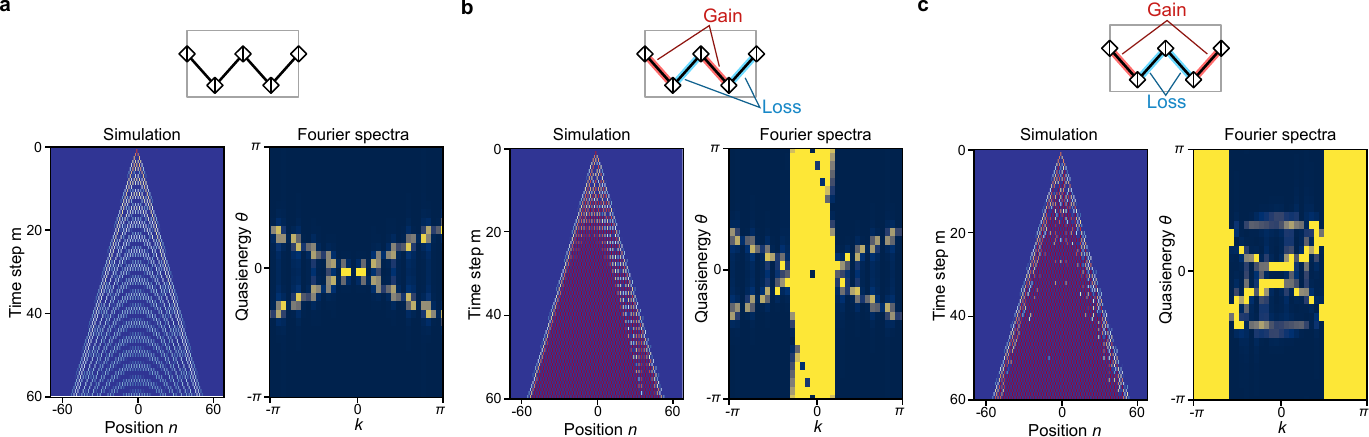}
  \caption{\textbf{Optical energy evolution and Fourier spectra for different gain/loss and phase distributions.} \textbf{a}, No gain/loss ($g = 0$), $\phi = \pi$.  \textbf{b}, Gain/loss/gain/loss, $\phi = \pi$, $g = 0.4$. \textbf{c}, Gain/loss/loss/gain.  At $k = 0$, $\phi = \pi$, $g = 0.4$. In \textbf{a}--\textbf{c}, the simulated intensity evolutions are obtained with a single pulse injection in the long loop at time $m = 0$. In each subplot, the Fourier spectra is derived from Fourier transformation of the simulation results on the left. All the other parameters are the same as these in experiments.}
  \label{figs16}
\end{figure*}

\begin{figure*}
  \centering
  \includegraphics[width=0.85\textwidth]{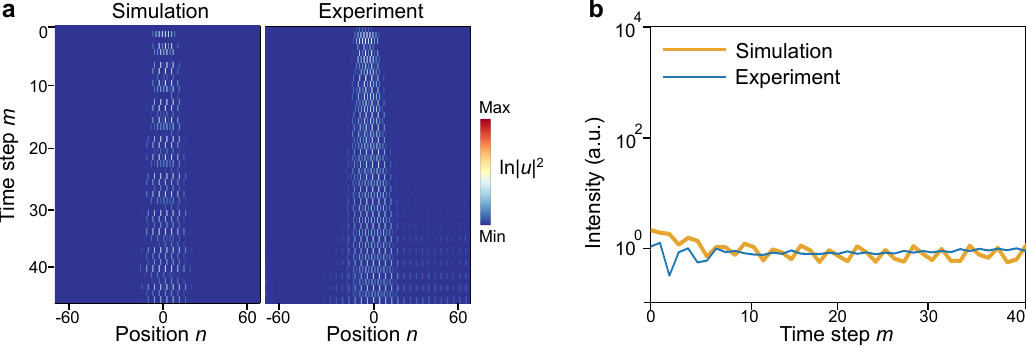}
  \caption{\textbf{Pulse propagation in the symmetry-broken regime.} \textbf{a}, Raw data showing the evolution of a wavepacket centered at $k = \pi$, with $g = 0.4055$, $\phi = \pi$, and uniform loss rate $\gamma = 0.223$.  \textbf{b}, Plot of the total intensity in the short loop versus $m$.}
  \label{figs6}
\end{figure*}

\begin{figure*}
  \centering
  \includegraphics[width=\textwidth]{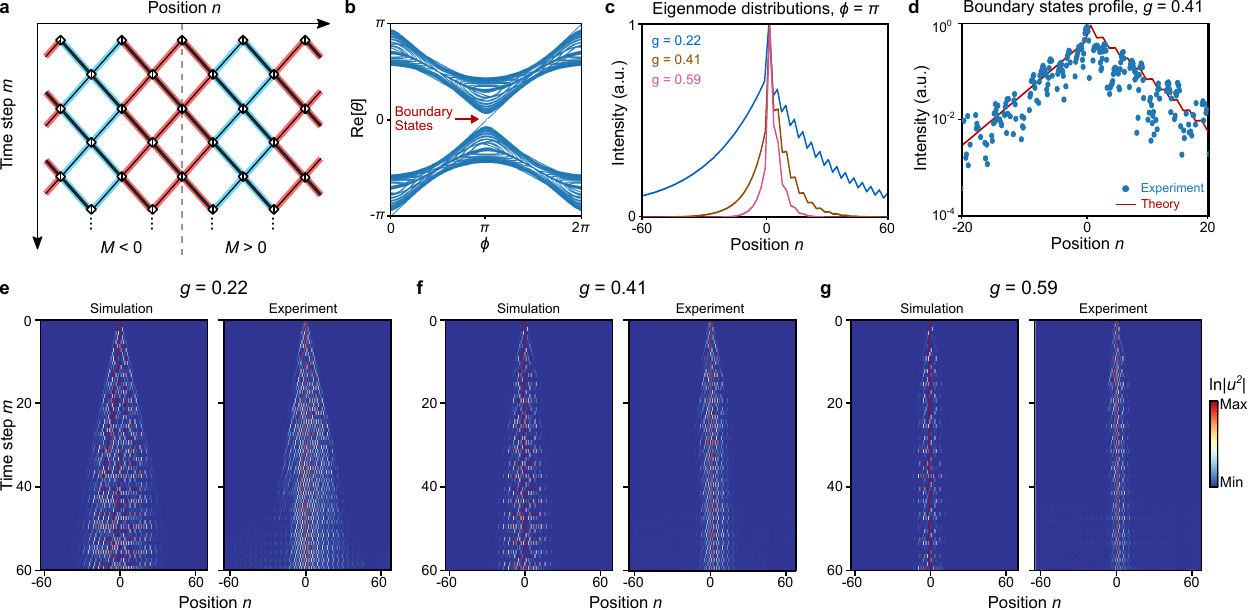}
  \caption{\textbf{Boundary states induced by gain and loss.}  \textbf{a}, Schematic of a synthetic lattice with two domains separated by a boundary (dashes).  The effective Dirac Hamiltonians have mass $M < 0$ and $M > 0$ in the left and right domains, respectively. Here, the color represents the psudospin of the Dirac quasiparticle, which is derived from the inner product between the eigenvector and the eigenvector of a massless Dirac Hamiltonian (the branch with positive group velocity).  \textbf{b}, Floquet band diagram for a finite sample of the lattice shown in \textbf{a}, with total size $N = 59$ and gain/loss level $g = 0.59$.  A chiral dispersion relation, corresponding to gain/loss induced boundary states, spans the gap.  \textbf{c}, Spatial distribution of $|u_{n}|^2+|v_{n}|^2$ for the mid-gap boundary state, calculated with different values of $g$.  \textbf{d}, Spatial profiles of $|u_{n}^m|^2+|v_{n}^m|^2$ for the theoretical boundary eigenstate for $g = 0.41$ (red line), and the corresponding experimental data time-averaged over $40 \le m \le 70$ (blue dots). In \textbf{c}--\textbf{d}, we normalize the maximum intensity for individual subplots to 1.  \textbf{e}--\textbf{g}, Time evolution of a pulse injected at $n = 0$, for $g = 0.22$, $0.41$, and $0.59$.  In simulations (left panels) and experimental data (right panels), the pulse spreads into a wavefunction localized at the boundary, with localization length following the trend shown in \textbf{c}.  In \textbf{c}--\textbf{g}, we set $\phi = \pi$.}
	\label{fig3supp}
\end{figure*}

\ignore{\begin{figure*}
  \centering
  \includegraphics[width=\textwidth]{FigureS11}
  \caption{\textbf{Behavior of boundary states.} \textbf{a}, Schematic of the lattice with a domain wall.  Additional gain is added to $F(0)$ to compensate for the imaginary component of the boundary state's quasienergy.  \textbf{b}, Floquet band diagrams for a finite sample of the lattice (with $N = 59$ sites and bulk gain/loss level $g = 0.59$), for different values of the compensating gain level $g' = 0.22, 0.41, 0.59$.  \textbf{c},  Distribution of $|u_{n}^m|^2$ for the boundary eigenstate at these values of $g'$.}
  \label{figs11}
\end{figure*}}

\begin{figure*}
  \centering
  \includegraphics[width=\textwidth]{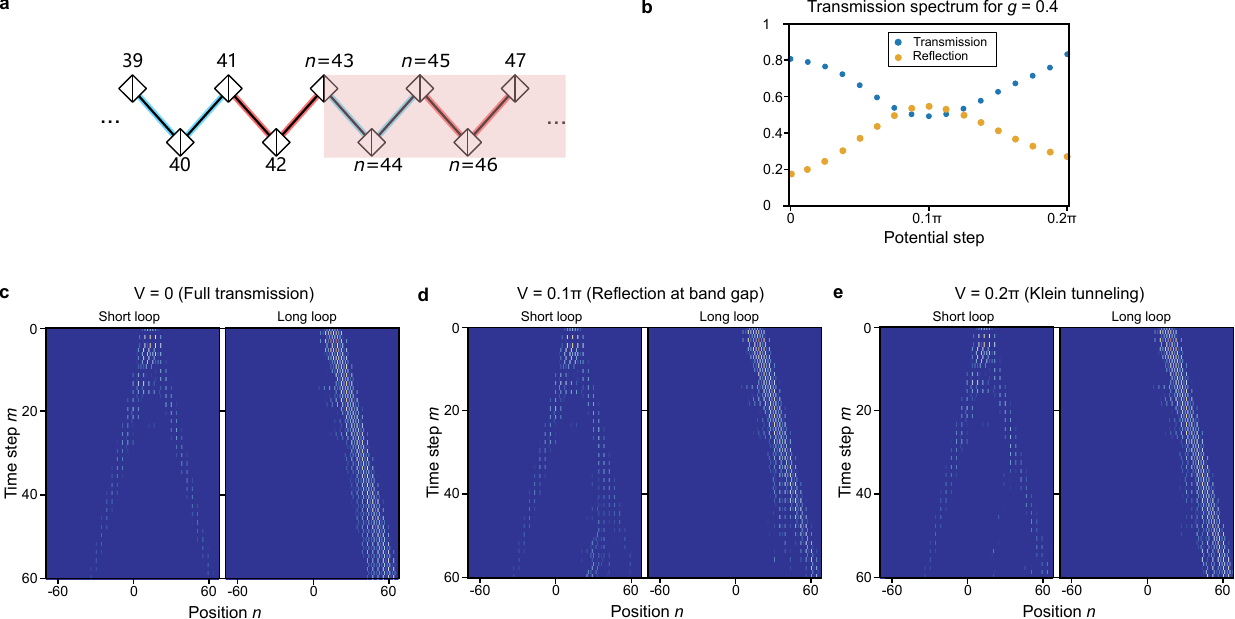}
  \caption{\textbf{Pulse propagation at different potential barrier heights.} \textbf{a}, Schematic of the lattice configuration near the interface. \textbf{b}, Transmission and reflection coefficient as a function of potential $V$. \textbf{c-e}, simulated intensity evolution as the potential barrier varies from $0$ to $0.2\pi$.  In \textbf{c}--\textbf{e}, we set $\phi = \pi$, $g = 0.4$ (which corresponds to $M \approx 0.08$), $k = 0.1\pi$ (which corresponds to $E \approx 0.03\pi$).}
  \label{figs14}
\end{figure*}

\begin{figure*}
  \centering
  \includegraphics[width=0.6\textwidth]{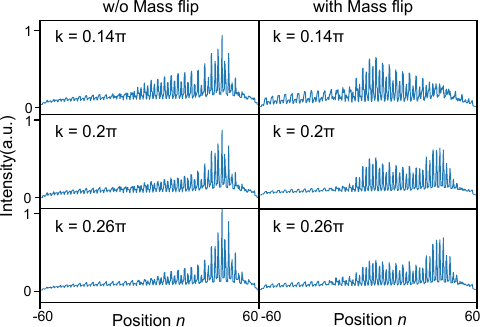}
  \caption{\textbf{Pulse profile at time step $m=42$.} Total intensity distribution with Gaussian pulse excitations of different momentum $k$. A temporal boundary is introduced at $m=10$ for the right subplot. For all figures, we set $\phi = \pi$, $g = 0.48$. }
  \label{figs20}
\end{figure*}

\clearpage

\pagebreak
\widetext
\begin{center}
\textbf{\large Supplementary Information for \\``Dirac mass induced by optical gain and loss''}
\end{center}

\setcounter{equation}{0}
\setcounter{figure}{0}
\setcounter{table}{0}
\setcounter{page}{1}

\renewcommand{\theequation}{S\arabic{equation}}
\renewcommand{\thefigure}{S\arabic{figure}}
\renewcommand{\thesection}{S\arabic{section}}

\section{Evolution equations for synthetic lattice}
\label{sec:evolution}

\begin{figure}[b]
  \centering
  \includegraphics[width=0.4\textwidth]{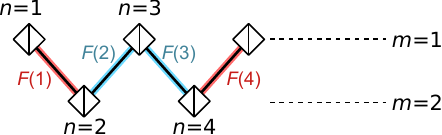}
  \caption{Schematic of the synthetic lattice.}
  \label{figs0}
\end{figure}

In this section, we discuss the derivation of the bandstructure for the non-Hermitian synthetic lattice.  This begins with the following evolution equation from the main text:
\begin{align}
  u_n^{m+1}&=\frac{1}{\sqrt{2}}(u_{n+1}^{m}+iv_{n+1}^{m}) \, e^{i\Phi(m)} \, F(n)
  \label{evol_orig_1}  \\
  v_n^{m+1}&=\frac{1}{\sqrt{2}}(iu_{n-1}^{m}+v_{n-1}^{m}) \, F(n-1),
  \label{evol_orig_2}
\end{align}
where
\begin{align}
  \Phi(m) &= (-1)^m\phi, \label{phim} \\
  F(n) &= \begin{cases} e^{g/2}, &\textrm{(gain)}\\
    e^{-g/2}, & \textrm{(loss)}.
  \end{cases}
\end{align}
The unit cell is shown in Fig.~\ref{figs0}.  This corresponds to the gain/loss distribution corresponding to Fig.~1\textbf{e} of the main text.

Evidently, the lattice has a temporal period of $\Delta m = 2$ and a spatial period of $\Delta n = 4$.  Hence, let us expand Eqs.~\eqref{evol_orig_1}--\eqref{evol_orig_2} into the following two-step equations valid for $n \equiv 1~(\textrm{mod}~4)$ and $m \equiv 1~(\textrm{mod}~2)$:
\begin{align}
  u_n^{m+2} &=
  \frac{1}{2} \left[u_{n+2}^{m}+iv_{n+2}^{m}
  + \left(- u_n^m + i v_n^m \right) e^{g+i\phi} \right] \label{step1}\\
  v_n^{m+2} &=
  \frac{1}{2} \left[(iu_{n-2}^{m}+v_{n-2}^{m}
    + \left(iu_n^m-v_n^m\right) e^{g-i\phi} \right]\\
  u_{n+2}^{m+2} &=
  \frac{1}{2} \left[u_{n+4}^{m}+iv_{n+4}^{m}
    + \left(-u_{n+2}^m+iv_{n+2}^m\right) e^{-g + i\phi} \right]\\
  v_{n+2}^{m+2} &=
  \frac{1}{2} \left[ iu_{n}^{m}+v_{n}^{m}
    + \left(u_{n+2}^m-v_{n+2}^m\right) e^{-g-i\phi} \right]. \label{step4}
\end{align}

We then take the following ansatz:
\begin{equation}
\begin{pmatrix}
u_n^m \\ v_n^m \\ u_{n+2}^m \\ v_{n+2}^m 
\end{pmatrix}
=
\begin{pmatrix}
U_0 \\ V_0 \\ U_1 \\ V_1
\end{pmatrix}
e^{-\frac{im\theta}{2}}e^{\frac{ikn}{4}},
\label{kansatz}
\end{equation}
where $\theta$ is the Floquet quasienergy and $k$ is the Bloch wavenumber.  The Floquet quasienergy plays a role similar to the energy in the more familiar Hamiltonian-based eigenvalue problems, except that $\mathrm{Re}[\theta]$ is a periodic variable \cite{Leykam2016}.  Substituting this into Eqs.~\eqref{step1}--\eqref{step4} yields the following eigenvalue problem:
\begin{equation}
\frac{1}{2}
  \begin{pmatrix}
-e^{g+i\phi} & ie^{g+i\phi} & e^{\frac{ik}{2}} & ie^{\frac{ik}{2}} \\
ie^{g-i\phi} & -e^{g-i\phi} & ie^{-\frac{ik}{2}} & e^{-\frac{ik}{2}} \\
e^{\frac{ik}{2}} & ie^{\frac{ik}{2}} & -e^{-g+i\phi} & ie^{-g+i\phi} \\
ie^{-\frac{ik}{2}} & e^{-\frac{ik}{2}} & ie^{-g-i\phi} & -e^{-g-i\phi} \\
\end{pmatrix}
\begin{pmatrix}
U_0 \\ V_0 \\ U_1 \\ V_1
\end{pmatrix}
=
e^{-i\theta}
\begin{pmatrix}
U_0 \\ V_0 \\ U_1 \\ V_1
\end{pmatrix}.
\label{Hamiltonian}
\end{equation}
By solving this numerically, we can plot the Floquet band diagrams shown in Fig.~1\textbf{c}--\textbf{e} of the main text.  For the gain/loss distribution that we focus on (Fig.~\ref{figs0}), the evolution matrix satisfy the semi-Hermiticity symmetries described in the Methods section of the main text.  We can check numerically that the Dirac-like eigenstates around $k = 0$, $\phi = \pi$ have the stated properties discussed in the main text---i.e., the energies are precisely real and the two relevant bands closest to $\theta = 0$ are orthogonal to each other.

In Extended Data Fig.~2, we plot Floquet band diagrams for different choices of gain/loss distribution, obtained using such evolution matrix calculations:

\begin{itemize}
\item In the Hermitian case, in addition to the Dirac point at $\theta = k = 0$, $\phi = \pi$, there is another Dirac point at $\theta = k = \pi$, $\phi = 0$, which does not play a significant role in the present work.

\item For the gain/loss/gain/loss configuration, similar to Fig.~1\textbf{d} of the main text.  The Dirac point turns into a pair of exceptionl points (EPs).  The band energies are non-real at $k = 0$, $\phi = \pi$, while the band energies close to $k =\pm \pi$, $\phi = \pi$ are real. If $g$ is further increased, all the band quasienergies become complex.

\item
  For the semi-Hermitian gain/loss/loss/gain configuration, which adds a real mass to the Dirac point at $k = 0$, $\phi = \pi$.  Note that the band energies are non-real for $k = \pm \pi$; this case is used as a comparison in Fig.~2\textbf{c},\textbf{f} of the main text.
\end{itemize}



\section{Effective Dirac Hamiltonian}
\label{sec:effective_hamiltonian}

In this section, we consider the effective Dirac Hamiltonian arising from the synthetic lattice's evolution equation, Eq.~\eqref{Hamiltonian}.  As noted in the Methods section of the main text, the semi-Hermitian symmetry
\begin{align}
  \Sigma_0 U \Sigma_0 &= (U^{-1})^\dag \label{pseudoHerm} \\ 
  \{ iU,\Sigma_3\Sigma_1 T \}&=0, \label{antiPT}
\end{align}
ensures that the evolution matrix has exactly real eigenvalues with pairwise-orthogonal eigenstates (in the unbroken-symmetry regime).  When $g = k = 0$ and $\phi = \pi$, there is a Dirac point at $\theta = 0$; for small but nonzero $g$, the Dirac point degeneracy is lifted, but the semi-Hermitian symmetry remains unbroken.  Hence, the bandstructure near $k = 0$, $\phi = \pi$ must be governed by an effective Dirac equation with nonzero mass.

We can use Eq.~\eqref{Hamiltonian} to derive the Dirac mass.  For $k = 0$ and $\phi = \pi$, the evolution matrix reduces to
\begin{equation}
  U_0(g) = 
\frac{1}{2}
  \begin{pmatrix}
e^{g} & -ie^{g} & 1 & i \\
-ie^{g} & e^{g} & i & 1 \\
1 & i & e^{-g} & -ie^{-g} \\
i & 1 & -ie^{-g} & e^{-g}
\end{pmatrix}.
\end{equation}
Take the ansatz
\begin{equation}
  |\psi_\pm\rangle = \begin{pmatrix}
    1 \\ \pm 1 \\ a \\ \pm a
  \end{pmatrix}.
\end{equation}
Plugging this into the evolution equation $U_0(g)|\psi_\pm\rangle = \exp(-i\theta) |\psi_\pm\rangle$, we obtain
\begin{align}
  \frac{1}{2} \left[e^g \mp i e^g + a + ia\right] &= e^{-i\theta}\\
  \frac{1}{2} \left[1 \pm i + e^{-g} a \mp i e^{-g}a\right]
  &= e^{-i\theta} a.
\end{align}
The two other rows of the eigenproblem give exactly the same equations.  Eliminating $a$, we obtain the following equation for $\theta$:
\begin{equation}
  e^{-2i\theta} - e^{-i\theta}(1\mp i)\cosh g \mp i = 0.
\end{equation}
This has quasienergy solutions
\begin{align}
  \theta &= \pm \left[\frac{\pi}{4}
    - \cos^{-1}\left(\frac{\cosh g}{\sqrt{2}}\right)\right] \\
  &= \pm \frac{g^2}{2} + O(g^4),
  \label{thetaresult}
\end{align}
which have the property of being purely real for small $g$ (specifically, $g < \cosh^{-1}\sqrt{2}$), and vanishing when $g = 0$.

To develop an approximate expression for the effective Dirac equation, let us take small deviations from the Dirac cone,
\begin{equation}
  k = \delta k, \quad \phi = \pi+\delta\phi, \quad
  \theta = \delta\theta,
\end{equation}
and retain only terms up to first order in $\delta k$, $\delta \phi$, and $\delta \theta$ in Eq.~\eqref{Hamiltonian}:
\begin{equation}
  \frac{1}{2}
  \begin{pmatrix}
  e^{g}(1+i\delta\phi) & -ie^{g}(1+i\delta\phi) & 1+\frac{i\delta k}{2} & i-\frac{\delta k}{2} \\
-ie^{g}(1-i\delta\phi) & e^{g}(1-i\delta\phi) & i+\frac{\delta k}{2} & 1-\frac{i\delta k}{2} \\
1+\frac{i\delta k}{2} & i-\frac{\delta k}{2} & e^{-g}(1+i\delta\phi) & -ie^{-g}(1+i\delta\phi) \\
i+\frac{\delta k}{2} & 1-\frac{i\delta k}{2} & -ie^{-g}(1-i\delta\phi) & e^{-g}(1-i\delta\phi) \\
\end{pmatrix}
  |\Psi\rangle
\approx (1-i\delta \theta) |\Psi\rangle.
\label{effecHamilt}
\end{equation}
Note that the evolution matrix on the left side of Eq.~\eqref{effecHamilt}, derived from a Taylor expansion, spoils the semi-Hermiticity symmetry \eqref{pseudoHerm}--\eqref{antiPT}.  Hence, our results for the Dirac parameters will only be approximations.  Now, let us take
\begin{equation}
  |\Psi\rangle = \begin{pmatrix}|\psi_0\rangle \\ |\psi_1\rangle \end{pmatrix},
\end{equation}
where $|\psi_0\rangle$ and $|\psi_1\rangle$ are two-component vectors.  We can rewrite Eq.~\eqref{effecHamilt} into
\begin{align}
  e^g \Big[(\sigma_0-i\sigma_1)+i\delta\phi (\sigma_2+\sigma_3)\Big]
  |\psi_0\rangle
  + \Big[(\sigma_0+i\sigma_1)
    -\frac{i\delta k}{2} (\sigma_2 - \sigma_3)\Big]\psi_1
  &\approx 2(1-i\delta \theta)|\psi_0\rangle \label{tmp1} \\
  \Big[(\sigma_0+i\sigma_1) - \frac{i\delta k}{2}(\sigma_2-\sigma_3)\Big]
  |\psi_0\rangle
  + e^{-g} \Big[(\sigma_0-i\sigma_1)+i\delta\phi(\sigma_2+\sigma_3)\Big]
  |\psi_1\rangle
  &\approx 2(1-i\delta \theta)|\psi_1\rangle. \label{tmp2}
\end{align}
Multiplying Eq.~\eqref{tmp1} by $\sigma_0-i\sigma_1+\frac{i\delta k}{2}(\sigma_2-\sigma_3)$, and again retaining only terms up to first order, we obtain:
\begin{equation}
  |\psi_1\rangle
  \approx \left[(1-i\delta\theta)(\sigma_0 - i \sigma_1) + ie^g\sigma_1
    +\frac{i\delta k}{2}(\sigma_2-\sigma_3)
    + i e^g \left(-\delta\phi+\frac{\delta k}{2}\right)\sigma_3 \right] |\psi_0\rangle.
\end{equation}
This can be substituted into Eq.~\eqref{tmp2} to eliminate $|\psi_1\rangle$, thus reducing the problem into
\begin{align}
  H_D |\psi_0\rangle
  &\approx \delta\theta \left(\sigma_0 + iM\sigma_1\right)|\psi_0\rangle, \\
  H_D &=
  M(g) \sigma_1
  + \frac{1}{2} \left[\left(1-\frac{e^{-g}}{2}\right) \delta k
    - e^{-g} \delta \phi \right] \sigma_2
  + \frac{1}{2} \left[\left(\frac{e^{g}}{2}-1\right)\delta k
    - e^{g}\delta\phi\right] \sigma_3, \label{Hdirac}\\
  M(g) &= \cosh g - 1.
  \label{Meq}
\end{align}
If we let
\begin{equation}
  \sigma_0 + iM\sigma_1 \approx \exp(iM\sigma_1),
\end{equation}
which is valid for small $g$ (i.e., small $M$), then
\begin{equation}
  H_D \, \big(e^{iM\sigma_1/2} |\psi'\rangle\big)
  \approx \delta \theta \, \big(e^{iM\sigma_1/2} |\psi'\rangle \big),
\end{equation}
to lowest order in $M$ and the other perturbative parameters ($\theta$, $\delta k$, and $\delta \theta$).

$H_D$ has the form of an anisotropic Dirac equation.  The Dirac mass is $M$, given by Eq.~\eqref{Meq}, is consistent with out previous result \eqref{thetaresult}:
\begin{equation}
  M = \frac{g^2}{2} + O(g^4).
\end{equation}
To determine the anisotropy factors, we can take the dispersion relation of Eq.~\eqref{Hdirac} when $M$ is negligible:
\begin{align}
  \delta\theta^2 &= \frac{1}{4}
  \left[\left(1-\frac{e^{-g}}{2}\right) \delta k
    - e^{-g} \delta \phi \right]^2
  + \frac{1}{4} \left[\left(\frac{e^{g}}{2}-1\right)\delta k
    - e^{g}\delta\phi\right]^2 \\
  &= \frac{\delta k^2}{8} + \frac{\delta\phi^2}{2} + \, O(g^3).
\end{align}
The $\delta k \delta\phi$ term is also of order $g^4$, and thus also neglected.  Hence, by appropriately rotating around $\sigma_1$, we can bring $H_D$ into the form
\begin{equation}
  H_D' \approx
  M(g) \sigma_1
  + \frac{1}{\sqrt{8}} \delta k \, \sigma_2
  - \frac{1}{\sqrt{2}} \delta \phi \, \sigma_3,
  \label{Hdirac2}
\end{equation}
where corrections to $M$ and the Dirac velocities are $O(g^3)$.

It is important to note that the wavenumber $k$ was defined in Eq.~\eqref{kansatz} such that advancing by one unit cell (which consists of 4 sites) advances the phase by $\exp(ik)$.  Likewise, $\theta$ is defined such that two time steps advances the phase by $\exp(i\theta)$.  Hence, the Dirac velocity along $k$, expressed in terms of sites per time step, is
\begin{equation}
  v_D = \frac{1}{\sqrt{2}}.
\end{equation}

In Extended Data Fig.~3, we plot the Floquet band diagrams in the $\phi$ and $k$ directions, for different values of $g$.  For $g = 0$ and small nonzero values of $g$, the quasienergies calculated directly from Eq.~\eqref{Hamiltonian} are well approximated by the Dirac equation \eqref{Hdirac2}.

The above derivation above applies to the unit cell of configuration gain/loss/loss/gain shown in Fig.~\ref{figs0}.  In the main text, we discuss an alternative lattice whose unit cell is gain/gain/loss/loss.  Repeating the above derivation yields a mass $-M$.

\section{Scattering from a potential step}

\begin{figure}
\centering
\includegraphics[width=\textwidth]{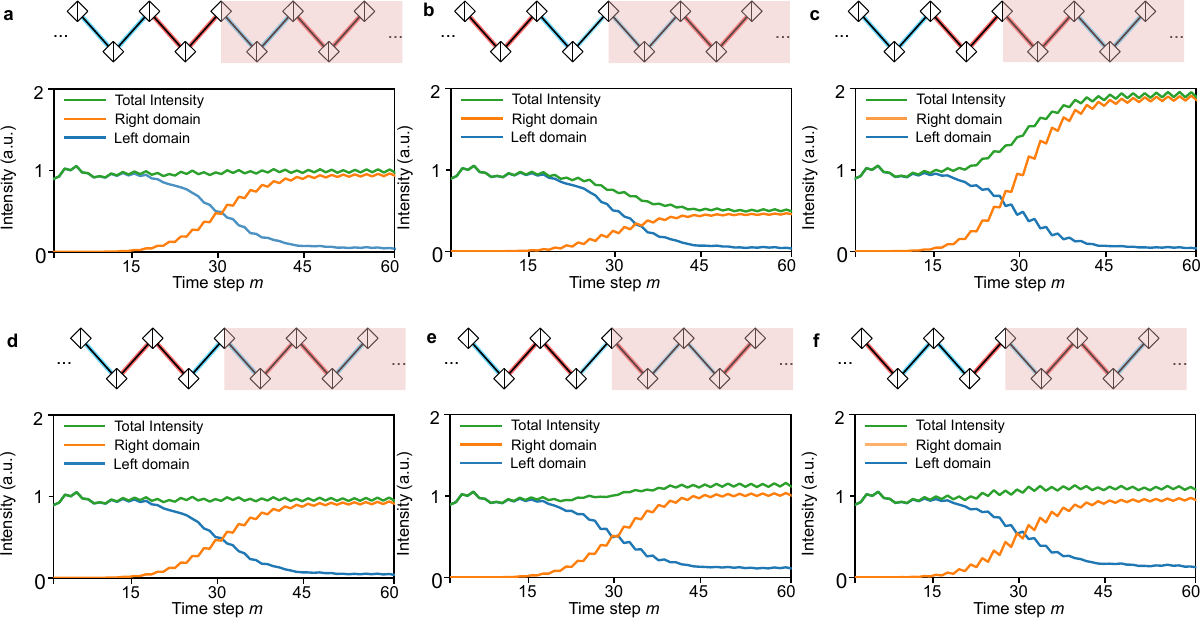}
\caption{Optical energy evolution for different types of heterojunctions. On the top of each subplot shows the unit cell configurations at two sides of the interface. In all the plots, we set $\phi = \pi$, $g = 0.3$, $k = 0.1\pi$, $V = 0.2\pi$.}
\label{fig:tunnelstep}
\end{figure}

In Fig.~4\textbf{d}--\textbf{e} of the main text, we plot the intensity evolution for scattering at heterojunctions between two uniform lattice domains.  The results show that the Dirac quasiparticle crosses the interface with either perfect transmission, amplification, or damping, depending on the configuration of the interface.

As noted in the main text, since the bulk domains have unbroken semi-Hermiticity, the observation of flux nonconservation implies that the interface itself acts as a source or sink.  This is verified in Fig.~\ref{fig:tunnelstep}, where we plot the evolution of the total intensity inside the lattice, as well as the intensity summed over the left and right sides of the interface.  Before and after the incident wavepacket strikes the interface (around $m = 30$), the intensities are approximately constant.

\begin{figure}
\centering
\includegraphics[width=\textwidth]{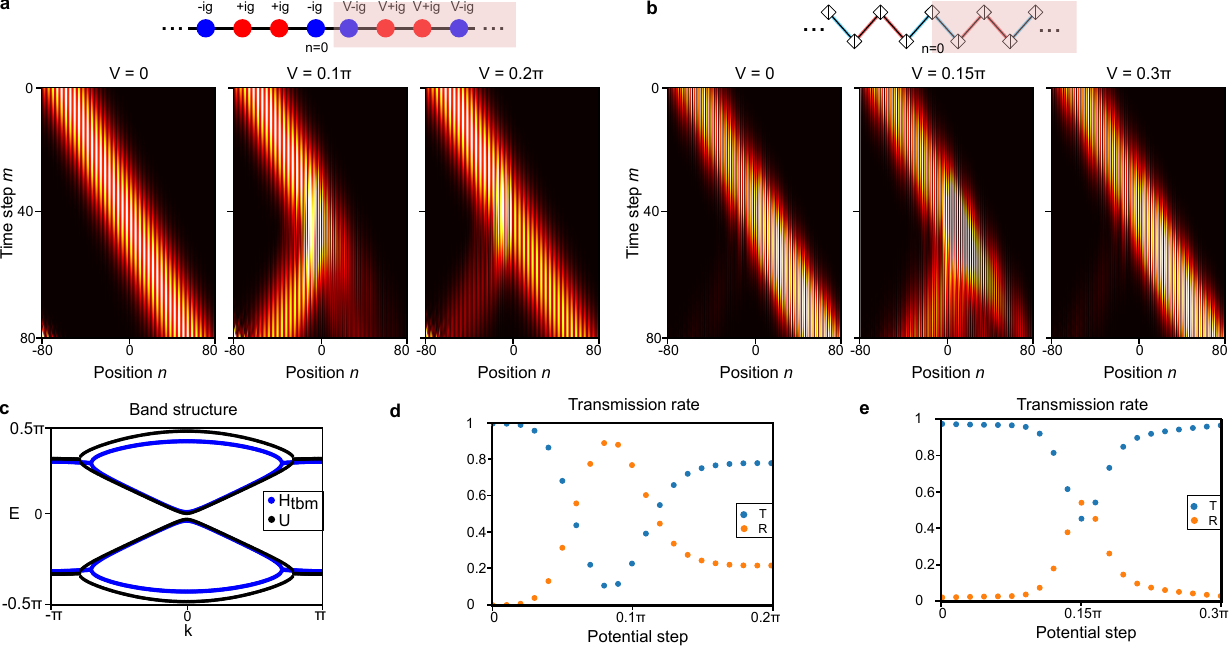}
\caption{Comparison of reflection and transmission behaviors between the Takata-Notomi model and the semi-Hermitian lattice model. \textbf{a}, Beam dynamics in a Takata-Notomi lattice, with Hamiltonian given by Eq.~\eqref{takataEquation}, with $g = 0.32$ and $t = -\frac{1}{\sqrt{2}}$.  Results are shown for three different values of the scalar potential barrier $V$.  \textbf{b}, Beam dynamics for the semi-Hermitian lattice with $\phi = \pi$ and $g = 0.32$. The lattice configurations are depicted above each subplot. \textbf{c}, Band structures of the Takata-Notomi lattice (TN, blue dots) and semi-Hermitian lattice (SH, black dots). Note that the lattices are chosen to have similar gaps and group velocities near $\mathrm{Re}(E) = 0$. \textbf{d}, Reflectance ($R$) and transmittance ($T$) versus $V$ for the Takata-Notomi lattice. \textbf{e}, Reflectance ($R$) and transmittance ($T$) versus $V$ for the semi-Hermitian lattice.}
\label{fig:takatanotomi}
\end{figure}


It is useful to compare the reflection and transmission behavior of the semi-Hermitian Dirac quasiparticles to that of the quasiparticles in the Takata-Notomi model \cite{Takata2018}.  This is a non-Hermitian 1D model possessing a real bandstructure, but with bands that are not orthogonal and not describable by an effective Dirac equation in the manner of Sec.~\ref{sec:effective_hamiltonian}.  The Takata-Notomi model's momentum space Hamiltonian is
\begin{equation}
  H_{TN}(k) = 
 \begin{pmatrix}
ig & t & 0 & te^{-ik} \\
t & ig & t & 0 \\
0 & t & -ig & t \\
te^{ik} & 0 & t & -ig
\end{pmatrix},
\label{takataEquation}
\end{equation}
where $g$ is the gain/loss level and $t$ is a coupling constant. To apply an additional scalar potential, we add $V$ to the on-site entries of the lattice Hamiltonian.
 
Fig.~\ref{fig:takatanotomi} compares the behavior of the two types of lattices, with the same gain/loss level $g = 0.32$.  The simulated wavepacket evolution for the two models, with different values of $V$, is shown in Fig.~\ref{fig:takatanotomi}\textbf{a}--\textbf{b}.  In Fig.~\ref{fig:takatanotomi}\textbf{c}, we plot the band structures for the two lattices, which look superficially similar near $\mathrm{Re}(E) \sim 0$.  However, the scattering behavior is quite different.  For large potential barriers the reflectance of the Takata-Notomi model is not suppressed to zero (Fig.~\ref{fig:takatanotomi}\textbf{d}).  For the semi-Hermitian lattice, when the potential barrier is strong the reflectance approaches zero and the transmittance approaches unity (Fig.~\ref{fig:takatanotomi}\textbf{e}), consistent with the Klein tunneling of Hermitian Dirac particles.  The behavior for larger $g$ is also qualitatively similar.

\vskip 0.1in
\section{Scattering by a temporal boundary}

In this section, we discuss the temporal reflection of Dirac quasiparticles and the derivation of the transmission parameters in Fig.~4\textbf{e} of the main text. Unlike the case of Klein tunneling, as the reflected and transmitted wavefunction are now orthogonal to each other. We can calculate the transmission by considering the projection operator onto respective eigenvectors directly.

\begin{figure}
\centering
\includegraphics[width=\textwidth]{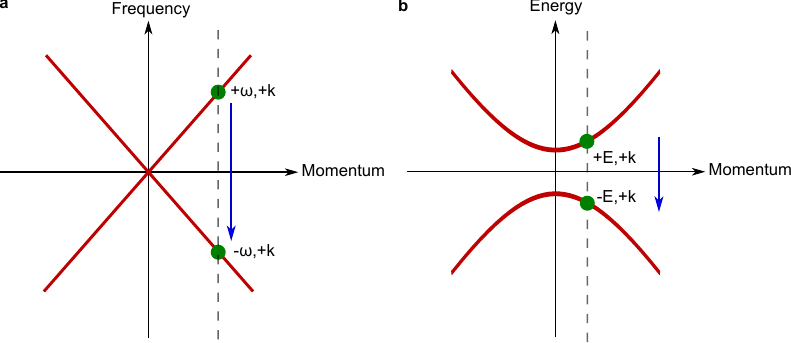}
\caption{Schematic of time-reflection for a scalar wave and a Dirac particle. \textbf{a}, For a scalar wave with dispersion $\omega = \pm ck$, time-reflection under conserved $k$ can be realized via the indicated vertical transition between positive and negative frequencies. \textbf{b}, For a massive Dirac particle, time-reflection can be achieved via a similar vertical transition, which maps to time-reversal in the $k\rightarrow 0$ limit.}
\label{figs14}
\end{figure}

Consider a massive Dirac equation of the form
\begin{equation}
(-i\vec{\alpha}\cdot\vec{\nabla}+\beta m)\psi = i\frac{\partial\psi}{\partial t}
\label{dirac_orig}
\end{equation}
where $\vec{\alpha}, \beta$ are the Dirac alpha matrices.  We consider plane-wave solutions of the form $\psi = \omega e^{-i(Et-\vec{p}\cdot\vec{x})}$, where $\omega$ is a 4-spinor composed of two 2-spinors,
\begin{equation}
\omega = \begin{bmatrix}
\phi \\
\chi
\end{bmatrix}
\end{equation}
Eq.~\eqref{dirac_orig} can then be simplified as
\begin{equation}
E\begin{bmatrix}
\phi \\
\chi
\end{bmatrix}
=\begin{pmatrix}
m\mathbb{I} & \vec{\sigma}\cdot\vec{p}\\
\vec{\sigma}\cdot\vec{p} & -m\mathbb{I}
\end{pmatrix}
\begin{bmatrix}
\phi \\
\chi
\end{bmatrix}.
\end{equation}
Solving the coupled equations results in a positive-energy solution of
\begin{equation}
\psi^{(+,+m)} = \sqrt{\frac{E+m}{2m}}\begin{bmatrix}
\phi \\
\frac{\vec{\sigma}\cdot\vec{p}}{E+m}\phi
\end{bmatrix}
e^{-i(Et-\vec{p}\cdot\vec{x})}.
\label{psiplus}
\end{equation}
We can undertake a similar derivation for a negative mass $-m$. The negative-energy solution, in this case, is
\begin{equation}
\psi^{(-,-m)}= \sqrt{\frac{-E+m}{2m}}\begin{bmatrix}
\phi \\
\frac{\vec{\sigma}\cdot\vec{p}}{E-m}\phi
\end{bmatrix}
e^{-i(Et-\vec{p}\cdot\vec{x})}.
\end{equation}
If we apply the ``time-reversal'' transformation $E \rightarrow -E$, we see that the negative-energy solution resembles the positive-energy solution in the $+m$ case, Eq.~\eqref{psiplus}, but with reversed momentum $-\vec{p}$:
\begin{equation}
\sqrt{\frac{E+m}{2m}}\begin{bmatrix}
\phi \\
\frac{\vec{\sigma}\cdot(-\vec{p})}{E+m}\phi
\end{bmatrix}
e^{-i(-Et-\vec{p}\cdot\vec{x})}.
\end{equation}
The probability of transitioning from $\psi^{(+,+m)}$ to $\psi^{(-,-m)}$ can be calculated as
\begin{align}
P &= |\langle \psi^{(-,-m)}|\psi^{(+,+m)}\rangle|^2\\ 
&= 1-\frac{| \vec{p} |^2}{(E+m)^2} 
\end{align}
As $\vec{p}\rightarrow 0$, $P\rightarrow 1$.  This derivation can also be adapted, with little change, to the 2D Dirac equation.  Thus, as shown in Fig.~\ref{figs14}, a momentum-conserving temporal boundary induces an efficient state transfer from the upper band to the lower band, which also reverses the group velocity.

\begin{figure}
\centering
\includegraphics[width=\textwidth]{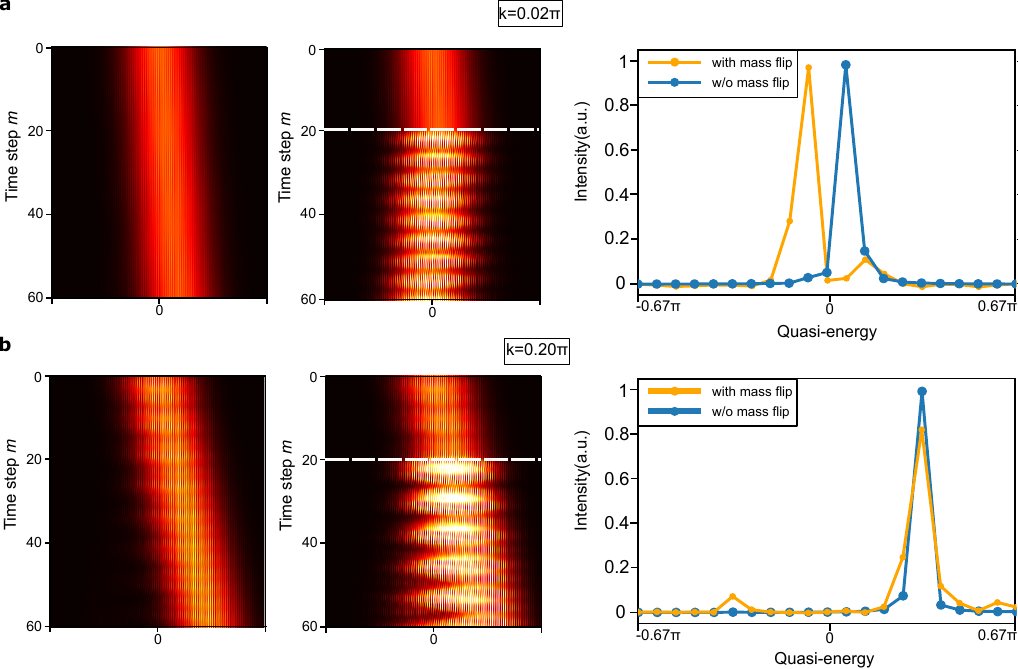}
\caption{Numerical simulations showing the scattering of a broad gaussian wavepacket by a temporal boundary. \textbf{a}, Beam dynamics at small momentum.  \textbf{b}, Beam dynamics at large momentum.  In each subplot, the panels show, from left to right, the evolution in the absence of a temporal boundary, the evolution with the temporal boundary, and the transmittance retrieved from a 2D FFT. For both subplots, we set $g = 0.48$ and $\phi=\pi$.}
\label{figs13}
\end{figure}

For the transmission parameters shown in Fig.~4\textbf{e} of the main text, starting from the two-band effective Dirac Hamiltonian Eq.~\eqref{Hdirac2}, if we set the eigenstates before the temporal boundary to be $(|\psi_1\rangle, |\psi_2\rangle)$ and after the boundary to be $(|\psi_1^{'}\rangle, |\psi_2^{'}\rangle)$, the transmission rate can then be calculated as $|\langle\psi_2^{'}|\psi_2\rangle|^2$, provided that the initial pulse is $|\psi_2\rangle$.

For the Floquet Hamiltonian \eqref{Hamiltonian}, there are four bands that are pairwise orthogonal, and an incident quasiparticle can be reflected into the other two non-Dirac bands. To account for this, we normalize the projected intensity with the total intensity associated with the two Dirac bands (i.e., the second and third band of Eq.~\eqref{Hamiltonian}). Provided the incident wavepacket is localized on the third band, $|\psi_3\rangle$, the transmittance is calculated as
\begin{equation}
  \frac{|\langle\psi_3^{'}| \psi_3\rangle|^2}{|\langle\psi_3^{'}|\psi_3\rangle|^2+|\langle\psi_2^{'}|\psi_3\rangle|^2}.
\end{equation}

Fig.~\ref{figs13} shows numerical simulations for the scattering of gaussian incident wavepackets much broader (and hence more well-localized in $k$-space) than in the experiment.  The results for small momentum (Fig.~\ref{figs13}\textbf{a}) and large momentum (Fig.~\ref{figs13}\textbf{b}) match the above predictions, as well as being consistent with the behaviors observed in our experiment (see main text).

\end{document}